# TAURID COMPLEX SMOKING GUN: DETECTION OF COMETARY ACTIVITY

**Ignacio Ferrín[(1)], Vincenzo Orofino[(2,3,4)]**

(1): Institute of Physics, Faculty of Exact and Natural Sciences, University of Antioquia, 0051000 Medellin, Colombia.
(2): Department of Mathematics and Physics "E. De Giorgi", University of Salento, 73100 Lecce Italy.
(3): INAF - Istituto Nazionale di Astrofisica, Section of Lecce.73100 Lecce Italy.
(4): INFN - Istituto Nazionale di Fisica Nucleare, Section of Lecce.73100 Lecce Italy.

Corresponding author: Vincenzo Orofino, Dipartimento di Matematica e Fisica "E. De Giorgi", Università del Salento, Via per Arnesano s.n., 73100 Lecce, Italy
(email: vincenzo.orofino@unisalento.it)

**Abstract**

Using the Secular Light Curve (SLC) formalism (Ferrín, 2010), we have catalogued 88 probable members of the Taurid Complex (TC). 51 of them have useful SLCs and 34 of these (67%) exhibit cometary activity. This high percentage of active asteroids gives support to the hypothesis of a catastrophe that took place during the Upper Paleolithic (Clube and Napier, 1984), when a large short-period comet, arriving in the inner Solar System from the Kuiper Belt, experienced, starting from 20 thousand years ago, a series of fragmentations that produced the present 2P/Encke comet, together with a large number of other members of the TC. The fragmentation of the progenitor body was facilitated by its heterogeneous structure (very similar to a rubble pile) and this also explains the current coexistence in the complex of fragments of different composition and origin. We have found that (2212) Hephaistos and 169P/NEAT are active and members of the TC with their own sub-group. Other components of the complex are groups of meteoroids, that often give rise to meteor showers when they enter the terrestrial atmosphere, and very probably also the two small asteroids that in 1908 and 2013 exploded in the terrestrial atmosphere over Tunguska and Chelyabinsk, respectively. What we see today of the TC are the remnants of a very varied and numerous complex of objects, characterized by an intense past of collisions with the Earth which may continue to represent a danger for our planet.

## 1. Introduction

According to the Whipple's (1967) original proposal, the unusual and odd comet 2P/Encke (hereinafter 2P) is the main source of a complex system of massive meteor showers, collectively called Taurids, and also the main contributor to the zodiacal cloud. Based on this association between 2P and all these showers, Clube and Napier (1984) developed the Taurid-Complex (TC) giant comet hypothesis. This hypothesis proposes that a giant comet, of ~100 km in diameter (comparable to that of a typical Kuiper Belt Object), fragmented 10-20 ky ago, producing a complex of dust, meteor streams and relatively large bodies (including 2P). Clube and Napier (1984) also concluded that the Tunguska event of June 30th, 1908, was related to a member of this group.

In the literature the date proposed for the cataclysmic event that gave origin to the TC is approximately constraint by dynamic considerations. It varies between 20,000 and 30,000 years BP, depending on the dynamic models (Clube & Napier, 1984; Asher & Clube, 1993; Seargent, 2017): much more than that and now the TC would have been already dispersed, having completely lost the current partial compactness; much less than that and the TC would be much more compact than present.

According to Clube and Napier (1984) the most recent glaciation on Earth which began around 22,000 BP, was closely related to the supposed catastrophic event, due to the influx of cosmic material produced by the fragmentation.

Asher and Clube (1993) further developed this idea concluding that many of these near-Earth objects were trapped in the 7:2 mean motion resonance with Jupiter which gave them orbital stability.

Tomko and Neslusan (2019) found that 17 meteor showers reported in the IAU-MDC list of all showers, were dynamically related to 2P, implying significant meteoroid activity. In Sect. 4.4 we will show that in 2020 thirty showers associated to the TC have been identified in the IAU Meteor Showers database. This is a remarkable number.

Tubiana et al. (2015) could not find an agreement between the spectra of the Maribo and Sutter´s Mill meteorites and 2P who have similar orbits. However, they cannot rule out a connection as the spectral differences may be caused by secondary alteration of the surfaces of the NEOs.

It is important to note that 2P should not be the only "traditional comet" originated by the fragmentation of a primeval parent body. By "traditional comet" we mean an object, with a carbonaceous/organic crust surrounding a mainly icy body, characterized by periodic or occasional emission of volatiles from the interior, through fractures in the crust. According to some authors (see Seargent, 2017, for a review), other three comets are members of the TC: D/1766 G1 (Helfenzrieder), 169P/NEAT, and P/2003 T12, with orbital parameters similar to those of the TC.

Two other interesting comets in this respect are C/1833 S1 (Dunlop) and C/1966 T1 (Rudnicki). The former is listed in the MPC database as parabolic with e = 1.0. However, Seargent (2017), reporting a former orbital analysis made by Shulhof in 1888, suggests that this comet could have an elliptic orbit. A small inclination (i=7.3°), a very short period of 3.5 years and its value of the longitude of perihelion, are all consistent with a TC membership, not with an Oort Cloud object. The confusion arises because there are only 15 observations of the object. Due to these uncertainties, this comet has not been included in our sample.

On the other hand comet Rudnicki was probably linked to the TC (Olsson-Steel, 1987; Ziolkowsky, 1988, 1990). In particular, Ziolkowsky (1990) found that this object - due to a recent planetary perturbation - has now a slightly hyperbolic orbit, but in the past it was a periodic comet. In any case, at present comet Rudnicki is not a member of the TC and for this reason it has not been considered in our analysis.

What the present work points out, is that a large fraction (67%) of probable members of the TC with useful photometric data, show low level cometary activity, and that these members are related photometrically and dynamically. It looks like they are pieces of a much larger object that disintegrated long time ago.

It is interesting to notice that, as discussed before, the TC has 4 recognized *bona fide* comets as members: 2P/Encke, 169P/NEAT, P/2003 T12 and D/1766 G1 (Helfenzrieder). So it would not be surprising if there were more additional members with lower activity. In fact, that is expected. Furthermore, since solar insolation peaks at perihelion, it is also expected that any activity of an asteroid will show up at or near perihelion. This is also expected and confirmed.

## 2. Membership

The first step of our investigation was to determine membership into the Complex. To that goal we applied the often adopted modified D criterion (Southworth and Hawkins, 1963; Steel, Asher and Clube, 1991; Asher, Clube and Steel, 1993a; 1993b) to more than 140 candidates proposed in 14 works (Davies, 1986; Napier, 2010; Spurny et al., 2017; Asher and Clube, 1993; Ziolkowski, 2002; Jenniskens and Jenniskens, 2006; Babadzhanov et al., 2008; Popescu et al., 2014; Porubcan, Kornos and Williams, 2006; Olech et al., 2016; Jopek, 2011; Dumitru et al., 2017; Tubiana et al., 2015; Seargent, 2017). Even if in some of the previous works the membership to the TC has been assessed with the same criterion, we decided in any case to repeat the test using the updated orbital parameters of the proposed objects.

According with the modified D criterion, a celestial body with orbital parameters *a* (semi-major axis), *e* (eccentricity) and *i* (inclination), is a member of a group of objects (in the sense that it has a similar orbit to the rest of the group, suggestive of a common origin), if:

$$\sqrt{\left(\frac{a-a_R}{3}\right)^2 + (e-e_R)^2 + \left(2\sin\frac{i-i_R}{2}\right)^2} \leq D_c \,.$$

Here $a_R$ = 2.1 au, $e_R$ = 0.82, and $i_R$ = 4.0° are reference values, universally adopted in the literature when the D criterion is applied to the TC. They represent the average of the orbital parameters of the various meteoroid streams belonging to the complex (Steel, Asher and Clube, 1991). $D_c$ is the cutoff value for the membership. Obviously it is a crucial parameter to assess whether an object belongs to the group. In this paper we use $D_c = 0.25$ in complete agreement with a long series of works on the TC (Olsson-Steel, 1987; Asher, Clube and Steel, 1993b; Babadzhanov, 2001; Babadzhanov, Williams and Kokhirova, 2008; Valsecchi, D'Abramo and Boattini, 2015).

This criterion is modified with respect the original one, proposed by Southworth and Hawkins (1963), which takes also into consideration the other angular parameter, the longitude of perihelion (ϖ). However, according to Asher, Clube and Steel (1993a; 1993b), *"the longitude term should not be included because, whilst appropriate for many (narrow) streams, the Taurids have been widely dispersed in longitude, predominantly by Jovian perturbations, and therefore a conventional longitude term in the D-criterion would have too large a contribution."*

It is worthwhile to note the condition $D < D_c$ defines obviously a sharp boundary between members and non-members of a group, a situation that very rarely occurs in nature. Actually, mainly due the uncertainties on the orbital parameters of the objects and on the ones used as reference, the membership has to be assessed using a probabilistic approach. In other words, for a given object, the larger $D$ is than $D_c$, the greater the probability that that object is not a member. This being the case, it is not reasonable to exclude from the TC an object i.e. with $D = 0.27$. On the contrary, for an object i.e. with $D = 0.33$ one can says with an excellent degree of reliability that it is not a TC member. For these reasons, we consider all the objects with *D* between 0.25 and, let say 0.30 (an upper limit suggested by Porubcan, Kornos and Williams, 2006), as probable members to the TC, with their membership to the complex that is more or less uncertain depending on the deviation of *D* from $D_c$.

Table 1 lists some orbital and photometric parameters of the 88 out of 141 objects that satisfy the modified *D* criterion. In Tables 2, 3 and 4, we report the orbital parameters, along with the *D* parameter, of the selected objects, grouped according to their activity status, or to the lack of photometric data. We note that, in the case of 2P *D* = 0.14, which results from: a = 2.22 au, e = 0.85, and i = 11.8°.

As shown by the *D* parameter listed in these tables, only two bodies of our sample have $0.25 \leq D \leq 0.30$, so that, according the above discussion, we consider them as probable members to the TC, while for the remaining 86 objects one can says with an excellent degree of reliability that they belong to the complex.

This is, for example, the case of the asteroid 2004 TG10, whose orbit, according to Jenniskens and Jenniskens (2006), is similar to those of the Northern Taurid stream (see Sect. 4.4), and is characterized by one of the lowest *D* values of all the sample (0.06). As discussed in the next section, we tested this object for cometary activity and it gave a positive result. We also measured the phase coefficient finding $\beta = 0.063$ very similar to the value found for comet 2P (0.066 – Ferrín, 2008). Only those two objects show such a large value of this coefficient, linking them in their surface scattering properties.

Our selection allowed us to define the limits of the distributions of the various orbital parameters, and compare them with the limits of other authors (Table 5). Of particular interest is the distribution of the longitude of perihelion, defined as:

$$\varpi = \Omega + \omega$$

where Ω is the longitude of the ascending node and ω is the argument of perihelion. The distribution of the parameter $\varpi$ (shown in Figure 1) deserves a special attention. In fact, the longitude of perihelion, for the reasons discussed above, does not contribute to the D criterion; however, for the various members of the complex, despite the perturbations, it is not reasonable to expect a random trend in their values of $\varpi$, but rather a certain tendency to place themselves preferentially close to a given reference value. In analogy with what was discussed before, this reference value is the average of the longitudes of perihelion of the various meteoroid streams belonging to the complex, given by $\varpi_R = 140°$ (Steel, Asher and Clube, 1991; Asher, Clube and Steel, 1993a). Comet 2P has instead $\varpi = 161°$, which interestingly coincides with the average longitude of perihelion of the whole sample (see Table 5).

Actually, we have found that the values of $\varpi$ of the sample are not randomly distributed, since the average of the absolute values of their differences with respect to $\varpi_R$ is equal to 52°, significantly different from 90°, which is the average value expected in case of a random distribution (Napier, 2010). This confirms the conclusion arising from the modified *D* criterion that the selected objects are actually members of the TC.

Going back to Table 5, the reported discrepancies between our data and those obtained by the other authors are most probably due to the different size of the analyzed samples. For example, Napier (2010) listed 19 members of the TC. In our work, of the 141 objects tested, 88 passed the *D* criterion. An interesting case concerns the lower limit of the longitude of perihelion, that in the present work is equal to 6°, significantly smaller than the one (64°) found by Napier (2010).

One might wonder if an object with such a low value of $\varpi$, and so far from the minimum value found by Napier (2010), can be considered a plausible TC member. We think so. For example, it would seem strange to exclude from the TC a body like (2101) Adonis, which has a *D* value among the lowest of the sample, only because its longitude of perihelion value ($\varpi = 34°$) is very far from the lower limit of the sample studied by Napier (2010). What can be said is that probably this body was expelled long ago from the original body or has experienced more intense perturbations so that its semi-major axis has undergone a great precession compared to the others.

In Figure 2 we plot the location of the TC members in the phase space e vs a. The TC occupies a small area of the diagram.

**Table 1.** Orbital and photometric properties of 88 TC candidates, taken from various sources, that satisfy the modified D criterion (see text). The third column lists the number of observations available for each object, as reported by the Minor Planet Center (MPC) database (Holman, 2018). The fourth column classifies the object, according to its activity status, as active (+), inactive (-) or with not enough observations (N). The fifth column reports the amplitude of the SLC ($A_{sec}$) in magnitudes, the sixth gives the duration of the active period ($\Delta t_A$) in days, the seventh the orbital period ($P_{orb}$) in years, the eighth the perihelion distance (q) in au, the ninth gives the visual absolute magnitude ($m_V(1,1,0)$), the tenth gives the visual absolute magnitude $H_v$ from MPC, and in last column the phase coefficient ($\beta$) in mag/°. The quantities $A_{sec}$, $\Delta t_A$, $m_V(1,1,0)$, and $\beta$ have been determined in the present work.

| | Object | MPC Nobs | Act. stat. | Asec [mag] | $\Delta t_A$ [d] | Porb [y] | q [au] | $m_V(1,1,0)$ | Hv MPC | β [mag/°] |
|---|---|---|---|---|---|---|---|---|---|---|
| 0a | 2P/Encke Source 1 | 5052 | + | -7.20 | 181 | 3.30 | 0.33 | 15.49 | ---- | 0.066 |
| 0b | 2P/Encke Source 2 | ---- | + | -3.00 | 815 | 3.30 | 4.09 | 15.49 | ---- | 0.066 |
| 1 | (2101) Adonis | 131 | + | -2.90 | 290 | 2.57 | 0.44 | 19.0 | 18.8 | ---- |
| 2 | (2201) Oljato | 957 | + | -0.79 | 462 | 3.21 | 0.62 | 14.92 | 15.2 | 0.0344 |
| 3 | (2212) Hephaistos | 2502 | + | -0.60 | 300 | 3.17 | 0.35 | 12.65 | 13.8 | 0.0375 |
| 4 | (4183) Cuno | 2086 | + | -0.70 | 120 | 2.79 | 0.73 | 13.60 | 14.4 | 0.0340 |
| 5 | (4197) 1982 TA | 835 | - | ---- | ---- | 3.48 | 0.52 | 14.11 | 16.6 | 0.0353 |
| 6 | (4341) Poseidon | 520 | - | ---- | ---- | 2.49 | 0.59 | 15.25 | 15.9 | 0.0367 |
| 7 | (4486) Mithra | 763 | + | -0.60 | 470 | 3.19 | 0.74 | 15.15 | 15.6 | 0.0335 |
| 8 | (5143) Heracles | 2731 | + | -0.75 | 140 | 2.48 | 0.42 | 13.00 | 14.0 | 0.0400 |
| 9 | (5731) Zeus | 563 | + | -0.50 | 220 | 3.41 | 0.78 | 15.00 | 15.6 | 0.0377 |
| 10 | (6063) Jason | 1166 | + | -0.80 | 490 | 3.29 | 0.52 | 15.75 | 15.9 | 0.0281 |
| 11 | (8201) 1994 AH2 | 751 | + | -0.40 | 150 | 4.04 | 0.74 | 15.42 | 15.7 | 0.0313 |
| 12 | (16960) 1998 QS52 | 1167 | + | -1.00 | 231 | 3.27 | 0.31 | 14.00 | 14.3 | 0.0314 |
| 13 | (17181) 1999 UM3 | 243 | + | -0.72 | 215 | 3.65 | 0.78 | 16.66 | 16.4 | 0.0224 |
| 14 | (30825) 1990 TG1 | 1206 | + | -0.44 | 210 | 3.81 | 0.78 | 14.43 | 14.7 | 0.0332 |
| 15 | (69230) Hermes | 1030 | + | -0.55 | 190 | 2.13 | 0.62 | 16.7 | 17.5 | 0.0376 |
| 16 | (85182) 1991 AQ | 351 | + | -0.50 | 250 | 3.31 | 0.50 | 16.95 | 17.1 | 0.0333 |
| 17 | (85713) 1999 SS49 | 542 | - | ---- | ---- | 2.67 | 0.79 | 15.27 | 15.6 | 0.0322 |
| 18 | (100004) | 345 | + | -0.30 | 100 | 4.19 | 0.78 | 15.5 | 16.3 | 0.0450 |
| 19 | (106538) | 355 | + | -0.50 | 320 | 3.78 | 0.57 | 15.80 | 16.2 | 0.0309 |
| 20 | (139359) 2001 ME1 | 697 | + | -0.45 | 140 | 4.27 | 0.34 | 15.75 | 16.4 | 0.045 |
| 21 | (153792) | 90 | + | -1.00 | 50 | 3.04 | 0.54 | 17.86 | 18.2 | 0.0397 |
| 22 | (154276) | 642 | + | -0.60 | 42 | 2.23 | 0.53 | 16.50 | 17.6 | 0.0450 |
| 23 | (162195) | 143 | + | -0.65 | 40 | 2.02 | 0.36 | 19.25 | 19.2 | 0.0250 |
| 24 | (162210) 1999 SM5 | 200 | - | ---- | ---- | 3.49 | 0.70 | 17.80 | 19.1 | 0.0392 |
| 25 | (162695) 2000 UL11 | 110 | - | ---- | ---- | 3.09 | 0.77 | 18.90 | 20.1 | 0.0464 |
| 26 | (189008) | 449 | + | -0.64 | 124 | 3.19 | 0.44 | 16.09 | 16.3 | 0.0321 |
| 27 | (192642) | 560 | + | -0.54 | 53 | 4.30 | 0.61 | 15.50 | 16.3 | 0.0475 |
| 28 | (217628) Lugh | 121 | N | ------ | ------ | 4.08 | 0.76 | ------ | 16.6 | -------- |
| 29 | (252091) 2000UP3 | 350 | + | -0.60 | 175 | 3.38 | 0.54 | 16.60 | 17.1 | 0.0355 |
| 30 | (269690) | 240 | - | ---- | ---- | 2.83 | 0.79 | 18.45 | 18.4 | 0.0263 |
| 31 | (285540) 2000 GU127 | 300 | - | ---- | ---- | 3.04 | 0.57 | 17.75 | 18.5 | 0.0392 |
| 32 | (297274) 1996 SK | 380 | - | ---- | ---- | 3.80 | 0.50 | 16.20 | 16.8 | 0.0333 |
| 33 | (306367) 5025 P-L | 111 | - | ---- | ---- | 4.03 | 0.65 | 15.39 | 15.6 | 0.0344 |
| 34 | (312942) 1995 EK1 | 300 | - | ---- | ---- | 3.40 | 0.51 | 16.70 | 17.3 | 0.0401 |
| 35 | (380455) 2003 UL3 | 84 | + | -0.80 | 230 | 3.36 | 0.45 | 17.50 | 17.9 | 0.0424 |
| 36 | (382395) 1990 SM | 85 | N | ---- | ---- | 3.06 | 0.50 | ---- | 16.2 | ---- |
| 37 | (405212) | 241 | + | -0.65 | 220 | 1.61 | 0.37 | 18.50 | 18.0 | 0.0168 |
| 38 | (408752) 1991 TB2 | 118 | N | ---- | ---- | 2.93 | 0.42 | ---- | 17.0 | ---- |
| 39 | (446791) 1998 SJ70 | 150 | N | ---- | ---- | 3.35 | 0.66 | ---- | 18.3 | --- |
| 40 | (452639) | 77 | N | ---- | ---- | 3.39 | 0.29 | ---- | 18.2 | ---- |
| 41 | (488453) 1994 XD | 250 | - | ---- | ---- | 3.35 | 0.66 | 18.25 | 19.1 | 0.0446 |
| 42 | (503941) 2003 UV11 | 916 | + | -0.60 | 70 | 2.56 | 0.34 | 18.50 | 19.5 | 0.0450 |
| 43 | 1991 BA | 7 | N | ---- | ---- | 3.17 | 0.71 | ---- | 28.6 | ---- |
| 44 | 1991 GO | 91 | N | ---- | ---- | 2.67 | 0.66 | ---- | 20.0 | ---- |
| 45 | 1995 CS | 14 | N | ---- | ---- | 2.70 | 0.44 | ---- | 25.5 | ---- |
| 46 | 1995 FF | 13 | N | ---- | ---- | 3.51 | 0.67 | ---- | 26.5 | ---- |
| 47 | 1997 GL3 | 200 | + | -0.45 | 30 | 3.44 | 0.49 | 18.45 | 19.1 | 0.0373 |
| 48 | 1998 VD31 | 184 | N | ---- | ---- | 4.32 | 0.51 | ---- | 19.4 | ---- |
| 49 | 1998 BY7 | 39 | N | ---- | ---- | 2.87 | 0.79 | ---- | 21.5 | ---- |
| 50 | 1999 SJ10 | 35 | N | ---- | ---- | 3.13 | 0.62 | ---- | 19.4 | ---- |
| 51 | 1999 TT16 | 189 | - | ---- | ---- | 3.17 | 0.73 | 18.75 | 19.8 | 0.0605 |
| 52 | 1999 VK12 | 16 | N | ---- | ---- | 3.33 | 0.50 | ---- | 23.7 | ---- |
| 53 | 1999 VR6 | 154 | + | -0.80 | 110 | 3.25 | 0.53 | 20.90 | 20.8 | 0.035 |
| 54 | 1999 XK136 | 100 | N | ---- | ---- | 3.67 | 0.71 | ---- | 20.3 | ---- |
| 55 | 2000 EU70 | 90 | - | ---- | ---- | 3.31 | 0.52 | 18.90 | 18.9 | 0.0157 |
| 56 | 2000 GW127 | 78 | N | ---- | ---- | 3.04 | 0.57 | 18.17 | 19.4 | 0.060 |
| 57 | 2000 VZ44 | 12 | N | ---- | ---- | 3.00 | 0.54 | ---- | 21.0 | ---- |
| 58 | 2000 XJ44 | 23 | N | ---- | ---- | 3.11 | 0.62 | ---- | 20.2 | ---- |
| 59 | 2001 CA21 | 9 | N | ---- | ---- | 3.06 | 0.36 | ---- | 18.6 | ---- |
| 60 | 2001 FA58 | 90 | - | ---- | ---- | 3.40 | 0.63 | 21.50 | 21.4 | 0.0129 |
| 61 | 2001 QE34 | 300 | + | -0.70 | 30 | 3.17 | 0.57 | 18.35 | 19.0 | 0.0427 |
| 62 | 2001 QJ96 | 76 | N | ---- | ---- | 2.01 | 0.32 | ---- | 22.1 | ---- |
| 63 | 2001 UX4 | 118 | N | ---- | ---- | 2.26 | 0.43 | ---- | 19.1 | ---- |
| 64 | 2001 QO142 | 39 | N | ---- | ---- | 3.11 | 0.54 | ---- | 19.3 | ---- |
| 65 | 2002 XM35 | 9 | N | ---- | ---- | 3.60 | 0.37 | ---- | 23.0 | ---- |
| 66 | 2003 SF | 68 | N | ---- | ---- | 3.18 | 0.48 | ---- | 19.8 | ---- |
| 67 | 2003 WP21 | 35 | N | ---- | ---- | 3.39 | 0.49 | ---- | 21.8 | ---- |
| 68 | 2004 TG10 | 95 | + | 0.85 | ---- | 3.34 | 0.31 | 18.75 | 19.4 | 0.0625 |
| 69 | 2005 NX39 | 72 | N | ---- | ---- | 3.88 | 0.31 | ---- | 19.7 | ---- |
| 70 | 2005 TB15 | 144 | N | ---- | ----- | 2.44 | 0.44 | ---- | 19.5 | ---- |
| 71 | 2005 TF50 | 48 | N | ---- | ---- | 3.43 | 0.30 | ---- | 20.3 | ---- |
| 72 | 2005 UR | 74 | N | ---- | ---- | 3.39 | 0.27 | ---- | 21.6 | ---- |
| 73 | 2006 SO198 | 26 | - | ---- | ---- | 2.83 | 0.26 | 19.52 | 23.9 | 0.050 |
| 74 | 2007 UL12 | 191 | N | ---- | ---- | 2.75 | 0.38 | ---- | 21.1 | ---- |
| 75 | 2007 RU17 | 390 | - | ---- | ---- | 2.91 | 0.35 | 17.55 | 18.1 | 0.0368 |
| 76 | 2010 TU149 | 101 | N | ---- | ---- | 3.27 | 0.38 | ---- | 20.7 | ---- |
| 77 | 2011 UD | 120 | + | -0.70 | 12 | 2.90 | 0.44 | 20.25 | 20.7 | 0.0375 |
| 78 | 2011 TC4 | 133 | - | ---- | ---- | 2.57 | 0.42 | 19.90 | 20.3 | 0.0344 |
| 79 | 2012 UR158 | 134 | N | --- | ----- | 3.35 | 0.32 | ---- | 20.7 | ---- |
| 80 | 2014 NK52 | 34 | N | ---- | ----- | 3.26 | 0.36 | ---- | 21.3 | ---- |
| 81 | 2015 TD144 | 122 | N | ---- | ---- | 2.78 | 0.48 | 22.5 | 22.6 | ---- |
| 82 | 2015 TX24 | 59 | N | ---- | ---- | 3.41 | 0.29 | ---- | 21.5 | ---- |
| 83 | 2015 VH66 | 45 | N | ---- | ---- | 3.44 | 0.35 | ---- | 20.1 | ---- |
| 84 | 2016 SL2 | 26 | N | ---- | ---- | 2.95 | 0.47 | ---- | 25.4 | ---- |
| 85 | 2016 VK | 36 | N | ---- | ---- | 2.38 | 0.39 | ---- | 22.4 | ---- |
| 86 | D/1766 G1 (Helfenzrieder) | 0 | + | ---- | ---- | 4.35 | 0.41 | ---- | ---- | ---- |
| 87 | 169P/NEAT | 1250 | + | -7.20 | 98 | 4.20 | 0.60 | 15.3 | ---- | 0.0348 |
| 88 | P/2003 T12 | 287 | + | ---- | ---- | 4.16 | 0.60 | 19 | ---- | ---- |
| | Averages | | | -0.87 ±0.30 | 191 ±26 | 3.15 ±.08 | 0.49 ±.02 | 16.62 ±0.37 | 18.80 ±0.04 | 0.035 ±0.006 |

**Table 2**. Orbital elements from the MPC of active objects of the TC. For comparison, the orbital parameters of 2P, which is the prototype of the TC active objects, are: a = 2.22 au, e = 0.85, i = 11.8° and ϖ =161°, corresponding to D = 0.14.

| a [au] | e | i [°] | ϖ [°] | Name | D |
|---|---|---|---|---|---|
| 1,87 | 0,76 | 1,3 | 34 | (2101) Adonis | 0,11 |
| 2,17 | 0,71 | 2,5 | 173 | (2201) Oljato | 0,12 |
| 2,16 | 0,84 | 11,5 | 237 | (2212) Hephaistos | 0,13 |
| 1,98 | 0,63 | 6,7 | 171 | (4183) Cuno | 0,20 |
| 2,19 | 0,66 | 3,0 | 251 | (4486) Mithra | 0,16 |
| 1,83 | 0,77 | 9,0 | 177 | (5143) Heracles | 0,14 |
| 2,26 | 0,65 | 11,4 | 139 | (5731) Zeus | 0,22 |
| 2,21 | 0,77 | 4,9 | 146 | (6063) Jason | 0,06 |
| 2,53 | 0,71 | 9,6 | 189 | (8201) | 0,20 |
| 2,20 | 0,86 | 17,5 | 143 | (16960) | 0,24 |
| 2,37 | 0,67 | 10,7 | 150 | (17181) | 0,21 |
| 2,44 | 0,68 | 8,7 | 243 | (30825) | 0,20 |
| 1,65 | 0,62 | 6,0 | 127 | (69230) Hermes | 0,25 |
| 2,22 | 0,78 | 3,1 | 223 | (85182) | 0,06 |
| 2,60 | 0,70 | 16,0 | 89 | (100004) | 0,29 |
| 2,42 | 0,76 | 10,3 | 205 | (106538) | 0,16 |
| 2,10 | 0,74 | 11,0 | 160 | (153792) | 0,15 |
| 1,70 | 0,69 | 8,7 | 133 | (154276) | 0,20 |
| 1,60 | 0,77 | 5,9 | 124 | (162195) | 0,18 |
| 2,17 | 0,79 | 8,1 | 64 | (189008) | 0,08 |
| 2,64 | 0,77 | 6,8 | 250 | (192642) | 0,19 |
| 1,37 | 0,73 | 5,0 | 121 | (405212) | 0,26 |
| 1,45 | 0,76 | 5,9 | 157 | (503941) 2003 UV11 | 0,23 |
| 2,19 | 0,76 | 8,5 | 147 | 1999 VR6 | 0,10 |
| 2,03 | 0,78 | 8,8 | 144 | 2011 UD | 0,10 |
| 2,24 | 0,80 | 14,6 | 168 | (380455) | 0,19 |
| 2,23 | 0,86 | 4,2 | 162 | 2004 TG10 | 0,06 |
| 2,66 | 0,85 | 7,9 | 255 | D/1766 G1 | 0,20 |
| 2,60 | 0,77 | 11,0 | 34 | 169P/NEAT | 0,21 |
| 2,59 | 0,77 | 11,0 | 35 | P/2003 T12 | 0,21 |
| 2,27 | 0,78 | 6,7 | 97 | 1997 GL3 | 0,08 |
| 2,63 | 0,87 | 5,9 | 27 | (139359) 2001 ME1 | 0,19 |
| 2,16 | 0,74 | 5,6 | 234 | 2001 QE34 | 0,09 |
| 2,25 | 0,76 | 9,4 | 314 | (252091) 2000 UP3 | 0,12 |
| 2,66 | 0,87 | 17,5 | 314 | Maximum Value | 0,29 |
| 1,37 | 0,62 | 1,3 | 27 | Minimum Value | 0,06 |
| 2,18 | 0,75 | 8,2 | 157 | Average | 0,16 |

**Table 3**. Orbital elements from the MPC of inactive asteroids of the TC.

| a [au] | e | i [°] | ϖ [°] | Name | D |
|---|---|---|---|---|---|
| 2,30 | 0,77 | 12,6 | 129 | (4197) Morpheus | 0,17 |
| 1,84 | 0,68 | 11,9 | 98 | (4341) Poseidon | 0,21 |
| 1,92 | 0,64 | 10,8 | 143 | (85713) | 0,22 |
| 2,00 | 0,79 | 3,6 | 98 | (269690) | 0,05 |
| 2,43 | 0,79 | 2,0 | 121 | (297274) | 0,12 |
| 2,53 | 0,74 | 3,8 | 139 | (306367) | 0,16 |
| 2,00 | 0,87 | 10,1 | 136 | 2006 SO198 | 0,12 |
| 2,04 | 0,83 | 9,0 | 147 | 2007 RU17 | 0,09 |
| 1,49 | 0,72 | 3,1 | 150 | 2011 TC4 | 0,23 |
| 2,16 | 0,66 | 2,0 | 117 | 1999 TT16 | 0,16 |
| 2,12 | 0,64 | 2,2 | 125 | (162695) 2000 UL11 | 0,19 |
| 2,30 | 0,69 | 5,2 | 287 | (162210) 1999 SM5 | 0,14 |
| 2,26 | 0,72 | 8,2 | 270 | 2001 FA58 | 0,13 |
| 2,22 | 0,77 | 13,0 | 59 | 2000 EU70 | 0,17 |
| 2,35 | 0,73 | 4,3 | 346 | (488453) 1994 XD | 0,12 |
| 2,26 | 0,78 | 9,1 | 292 | (312942) 1995 EK1 | 0,11 |
| 2,10 | 0,73 | 8,5 | 311 | (285540) 2000 GU127 | 0,12 |
| 2,53 | 0,87 | 13,0 | 346 | Maximum Value | 0,26 |
| 1,49 | 0,64 | 2,0 | 59 | Minimum Value | 0,05 |
| 2,14 | 0,74 | 7,0 | 175 | Average | 0,15 |

**Table 4**. Orbital elements from the MPC of asteroids of the TC with not enough data.

| a [au] | e | I [°] | ϖ [°] | Name | D |
|---|---|---|---|---|---|
| 2,55 | 0,70 | 4,0 | 133 | (217628) Lugh | 0,19 |
| 2,11 | 0,76 | 11,6 | 243 | (382395) | 0,15 |
| 2,05 | 0,79 | 7,9 | 131 | (408752) | 0,08 |
| 2,26 | 0,87 | 12,2 | 165 | (452639) | 0,16 |
| 2,16 | 0,67 | 2,1 | 191 | 1991 BA | 0,15 |
| 1,93 | 0,65 | 9,6 | 114 | 1991 GO | 0,20 |
| 1,94 | 0,77 | 2,6 | 28 | 1995 CS | 0,08 |
| 2,65 | 0,81 | 11,6 | 161 | 1998 VD31 | 0,23 |
| 2,23 | 0,77 | 9,5 | 152 | 1999 VK12 | 0,12 |
| 2,25 | 0,73 | 6,6 | 100 | 2000 GW127 | 0.11 |
| 1,59 | 0,80 | 5,9 | 101 | 2001 QJ96 | 0,17 |
| 1,72 | 0,75 | 8,9 | 156 | 2001 UX4 | 0,17 |
| 2,35 | 0,84 | 3,1 | 184 | 2002 XM35 | 0,09 |
| 2,16 | 0,78 | 5,6 | 110 | 2003 SF | 0,05 |
| 2,26 | 0,78 | 4,3 | 161 | 2003 WP21 | 0,07 |
| 2,47 | 0,87 | 13,9 | 160 | 2005 NX39 | 0,22 |
| 1,81 | 0,76 | 7,3 | 149 | 2005 TB15 | 0,13 |
| 2,27 | 0,87 | 10,7 | 160 | 2005 TF50 | 0,14 |
| 2,26 | 0,88 | 7,0 | 162 | 2005 UR | 0,10 |
| 1,97 | 0,81 | 4,2 | 163 | 2007 UL12 | 0,04 |
| 2,20 | 0,83 | 2,0 | 152 | 2010 TU149 | 0,05 |
| 2,24 | 0,86 | 3,2 | 166 | 2012 UR158 | 0,06 |
| 2,20 | 0,84 | 2,5 | 163 | 2014 NK52 | 0,05 |
| 1,97 | 0,76 | 1,6 | 130 | 2015 TD144 | 0,09 |
| 2,27 | 0,87 | 6,0 | 160 | 2015 TX24 | 0,08 |
| 2,28 | 0,85 | 7,4 | 165 | 2015VH66 | 0,09 |
| 1,97 | 0,76 | 1,6 | 130 | 2016 SL2 | 0,09 |
| 1,78 | 0,78 | 5,9 | 166 | 2016 VK | 0,12 |
| 2,32 | 0,71 | 0,6 | 109 | 1995 FF | 0,14 |
| 2,13 | 0,75 | 5,5 | 102 | 2001 QO142 | 0,08 |
| 2,02 | 0,61 | 3,3 | 213 | 1998 BY7 | 0,22 |
| 2,24 | 0,71 | 7,3 | 268 | (446791) 1998 SJ70 | 0,14 |
| 2,38 | 0,70 | 2,7 | 6 | 1999 XK136 | 0,15 |
| 2,06 | 0,74 | 5,3 | 137 | 2000 VZ44 | 0,09 |
| 2,14 | 0,71 | 6,9 | 261 | 1999 SJ10 | 0,12 |
| 1,66 | 0,78 | 5,0 | 265 | 2001 CA21 | 0,15 |
| 2,13 | 0,71 | 10,6 | 339 | 2000 XJ44 | 0,16 |
| 2,65 | 0,88 | 13,9 | 339 | Maximum Value | 0,23 |
| 1,59 | 0,61 | 0,6 | 6 | Minimum Value | 0,04 |
| 2,13 | 0,77 | 6,1 | 158 | Average | 0,12 |

**Table 5.** Variation ranges of the orbital parameters of the TC members, obtained in the present work and by other authors; in the last line our average values of the same parameters are also reported.

| Author | a [au] | e | i [°] | $\varpi$ [°] | q[au] |
|---|---|---|---|---|---|
| Napier (2010) | 1.83<a<2.64 | 0.64<e<0.83 | 2.5<i<12.2 | $64<\varpi<251$ | 0.0 < q < 1.0 |
| Clark et al. (2019) | 2.23<a<2.27 | ---- | ---- | $145<\varpi<165$ | ---- |
| Spurny et al. (2017) | --- | --- | 4.5<i<4.6 | ---- | 0.25<q<0.45 |
| This work (limits) | 1.37<a<2.66 | 0.61<e<0.88 | 0.6<i<17.5 | $6<\varpi<346$ | 0.26<q<0.80 |
| This work (averages) | <a>=2.15±0.28 | <e>=0.76±0.07 | <i>=7.1±3.7 | $\langle\varpi\rangle=161\pm70$ | <q>=0.51±0.15 |

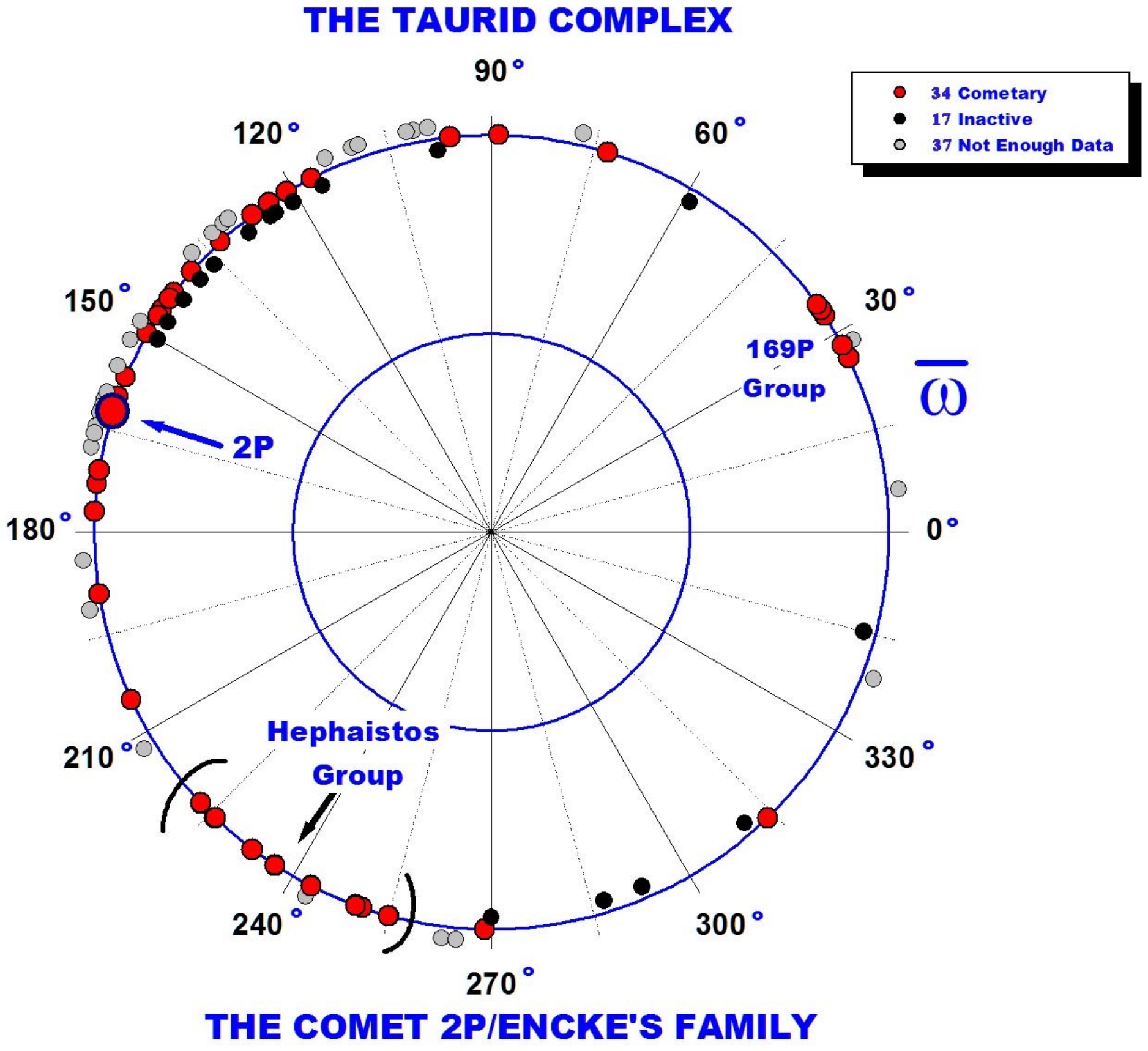

**Figure 1.** Polar diagram showing the distribution of the $\varpi$ parameters of TC members. The objects are placed at slightly different distances to the center, to be able to see the different groups distinctly.

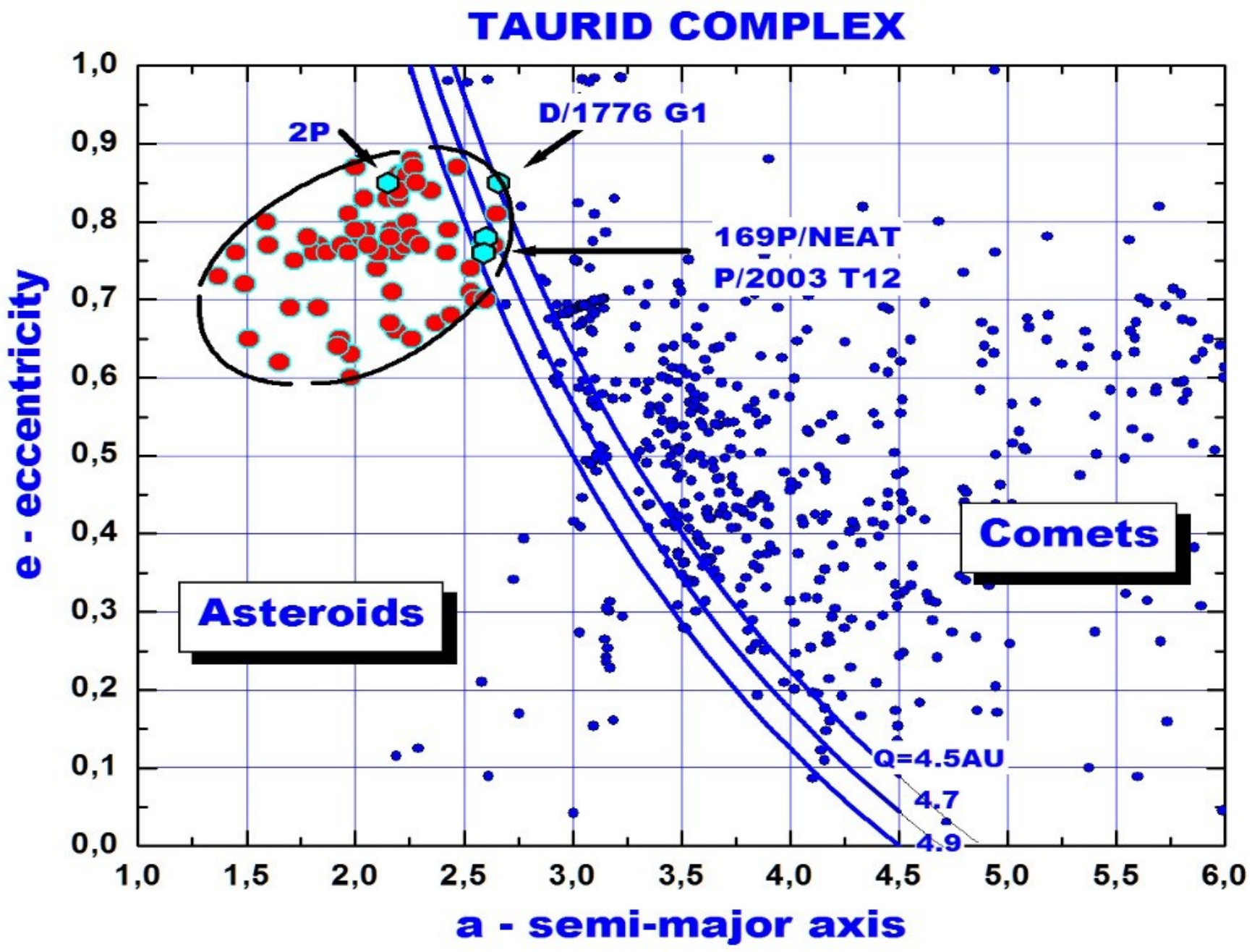

**Figure 2.** Eccentricity (e) vs semi-major axis (a) for members of the TC. The curved lines labeled with three different values of the aphelion distance (Q) at the bottom, separate comets from asteroids. The list of comets has been taken from the MPC. The TC is located in a reduced ellipsoidal space of the diagram. Four comets members of the TC have been labeled: 2P, D/1776 G1, P/2003 T12, and 169P/NEAT. Although 67% of the TC objects are active (as discussed in the text in Section 3), they lie in a region of the diagram populated by asteroids.

## 3. Search for activity and results

In order to search for activity, we used the Secular Light Curve (SLC) formalism (Ferrín, 2010) already successfully applied to study the cometary activity of various objects (Ferrín, 2014; Ferrín et al., 2017, 2018). For the convenience of the reader, some details about this technique are reported in Appendix 1. The SLC formalism produces a plot of absolute magnitude $m_v(1,1,0)$ vs $(t-T_q)$, where $m_v(1,1,0)$ is the magnitude at $\Delta$=1 au from the Earth, R=1 au from the Sun and phase angle $\alpha = 0°$, and $T_q$ is the time of perihelion. The absolute and the observed $m_v(\Delta, R, \alpha)$ magnitudes are related by:

$$m_v(1, 1, 0) = m_v(\Delta,R,\alpha) - 5 \log (\Delta \cdot R) - \beta \cdot \alpha$$

where $\beta$ is the phase coefficient. We used a linear law because the data fits well that law (example in Figure 5), and because the maximum values of the phase angle are not large.

We neglected all the observations with $\alpha < 5°$ because they produce an enhancement due to the opposition effect that creates a false positive.

Before concluding that an object is active, it is a good exercise to get acquainted with some negative results, shown in Figure 3.

In Figure 4 we report the SLC's of six active members of the TC, while Figures 5 and 6 show the phase plot of (2201) Oljato and its SLC, respectively. The SLC's of all the other active asteroids of the complex are presented in the Appendix 2. Finally, the SLC's of the two comets of the TC, 2P and 169P/NEAT, are shown in the next sections.

As reported in Table 1, 34 of 51 (equal to 67%) probable members of the TC with useful photometric data, show cometary activity. For comparison, Ferrín et al. (2017) made a similar study of 165 members of the Themis family and found that only 15% show signs of activity. So the large fraction of active bodies in the TC is highly indicative: this is the smoking gun of the complex.

Table 1 also shows that the active asteroids exhibit their cometary activity for a mean period of 191 days, which represents a duty cycle of 17%.

As far as the activity of asteroid Oljato is concerned, it is important to note that in four apparitions the object has shown in proximity of perihelion relatively low level cometary activity, which is the best evidence of ongoing sublimation, also suggested by hint of gaseous activity reported by McFadden et al. (1993) and A'Hearn et al. (1995).

Figure 5 shows the phase plot of Oljato to illustrate an important aspect of our reduction: the absolute magnitude, a fundamental parameter in this context, is determined using the SLC and the phase plot, both of which have to agree on $m_V(1,1,0)$. This procedure is very different from common determinations of the absolute magnitude found in the literature based on a few observations. This is the reason why we believe that our determinations of $m_V(1,1,0)$ is more reliable than other literature determinations, because the two plot are in different phase spaces, and in particular the SLC covers from aphelion ($R = -Q$) to aphelion ($R = +Q$), giving the whole picture of the absolute magnitude in the orbit.

The orbital elements of active (+) and inactive (–) asteroids of the TC are listed in Tables 2 and 3, while Table 4 reports the elements of asteroids with not enough data available (N). Note that the average values of the orbital parameters a, e and i do not show substantial differences between the active and non-active objects, as evidenced also by the average value of D. This means that the active objects do not have significantly different orbits from those of the objects not active.

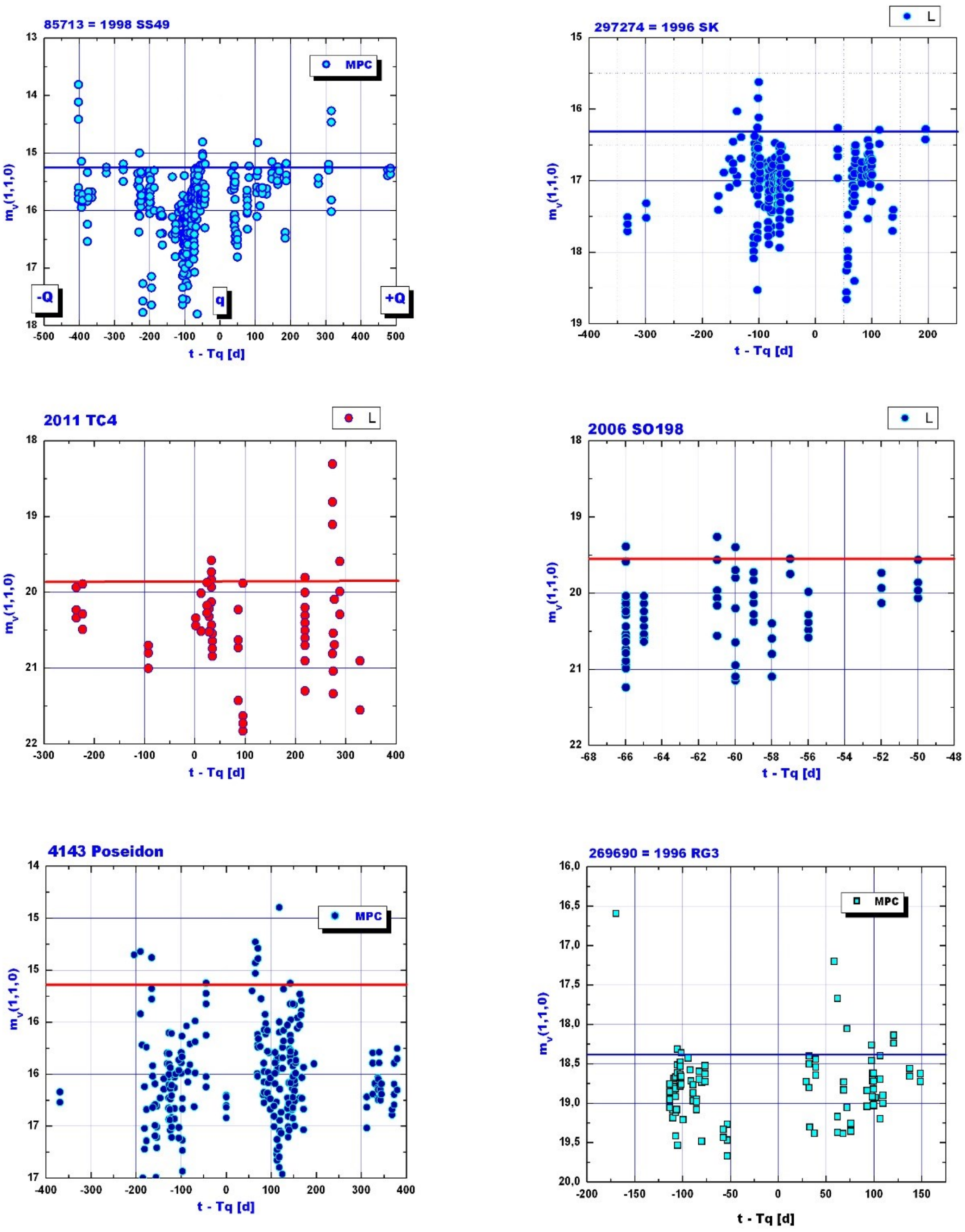

**Figure 3.** Examples of asteroids which do not show cometary activity.

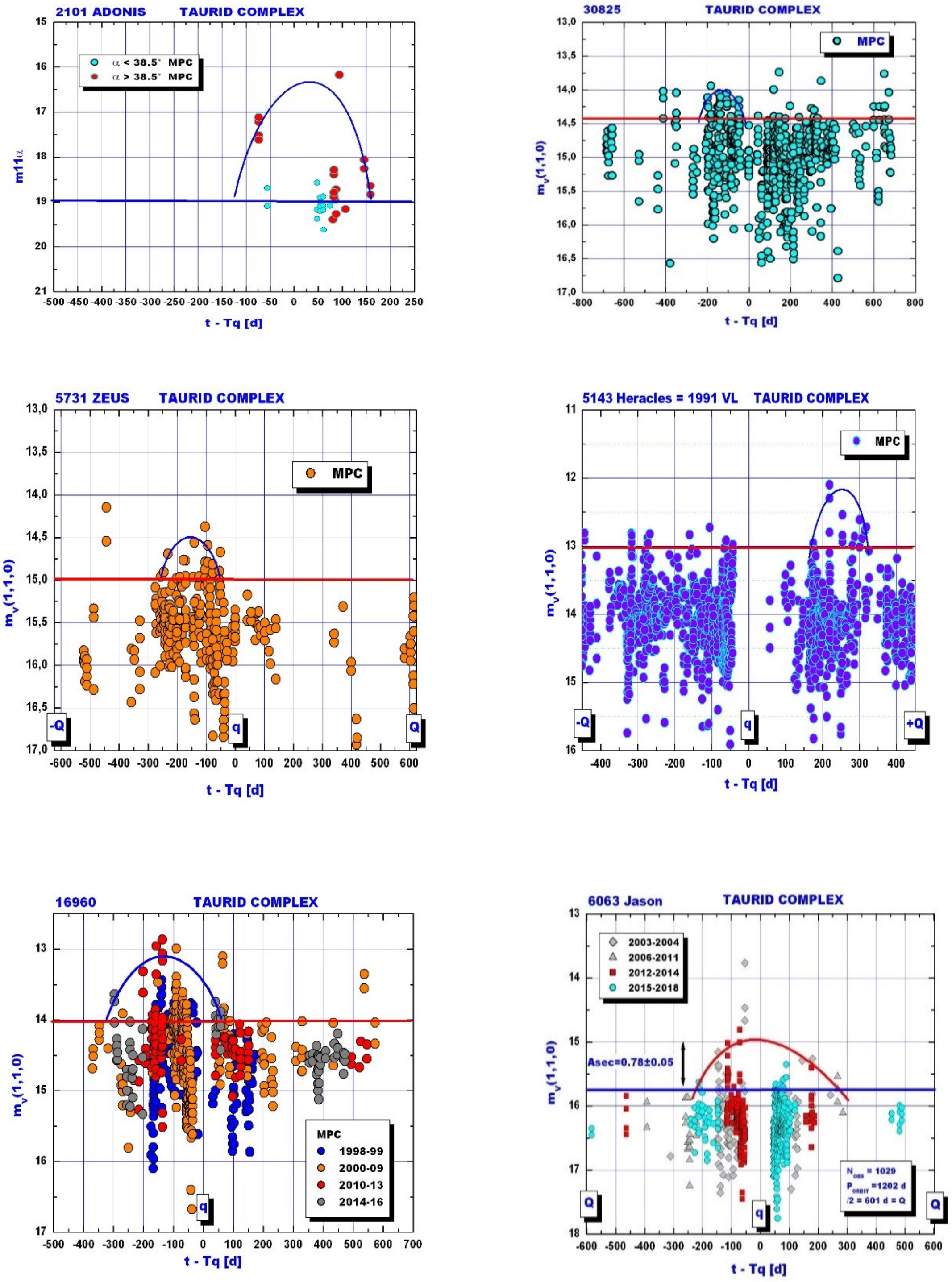

**Figure 4**. SLC's of six asteroids of the TC that exhibit low level cometary activity. Others appear in Appendix 2. Asteroid (16960) shows activity in five different apparitions, while (6063) Jason in three.

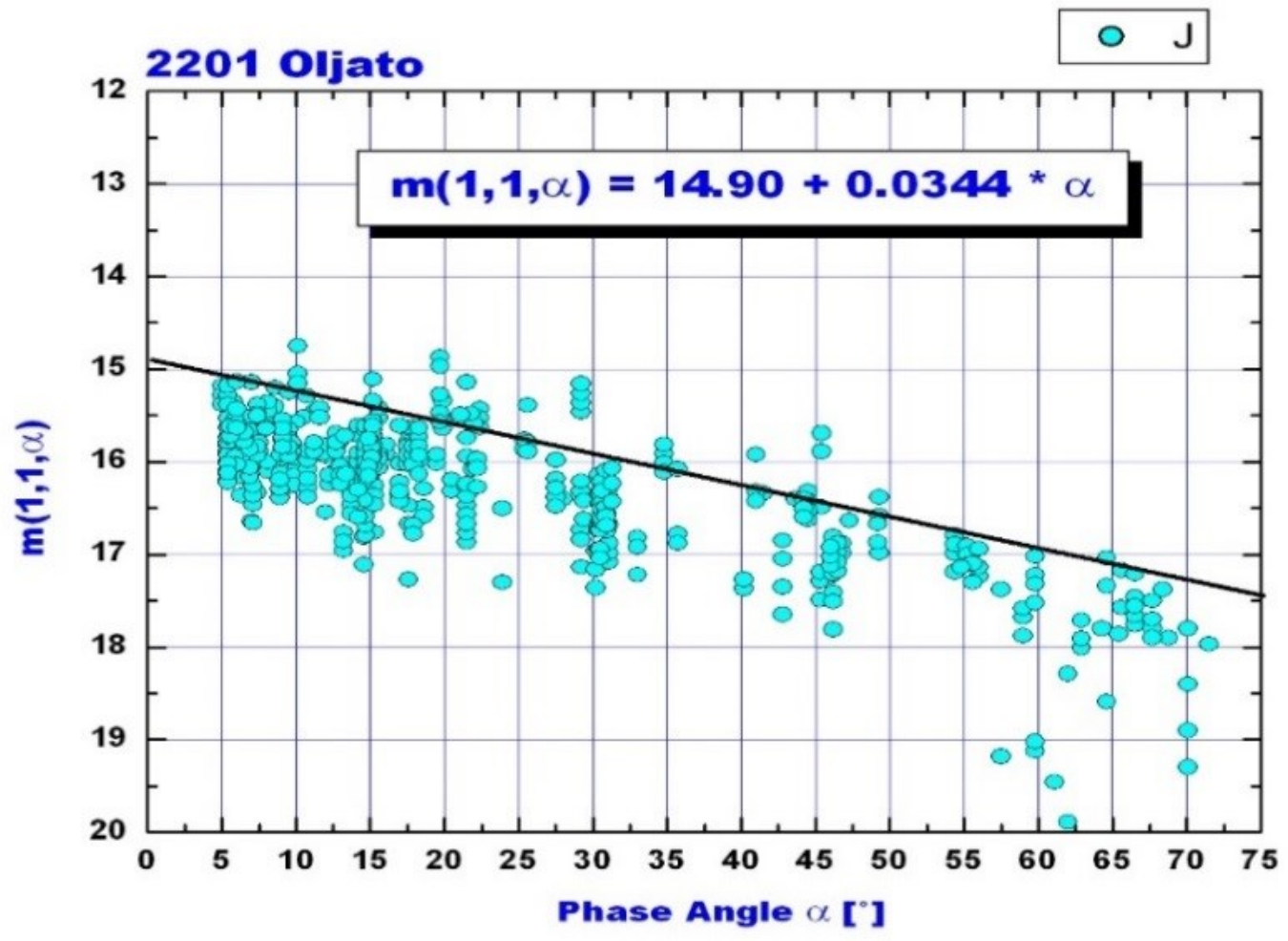

**Figure 5**. The phase plot of (2201) Oljato. This figure shows the close relationship between the phase and SLC plots. The phase plot of Oljato is well behaved and a phase relationship can be derived where the extrapolated value at $\alpha = 0°$ gives the absolute magnitude. The data with $\alpha < 5°$ has been removed because the opposition effect would produce false positives in the SLC. Notice that the data fits a linear law up to 70°, so a more sophisticated law is not needed.

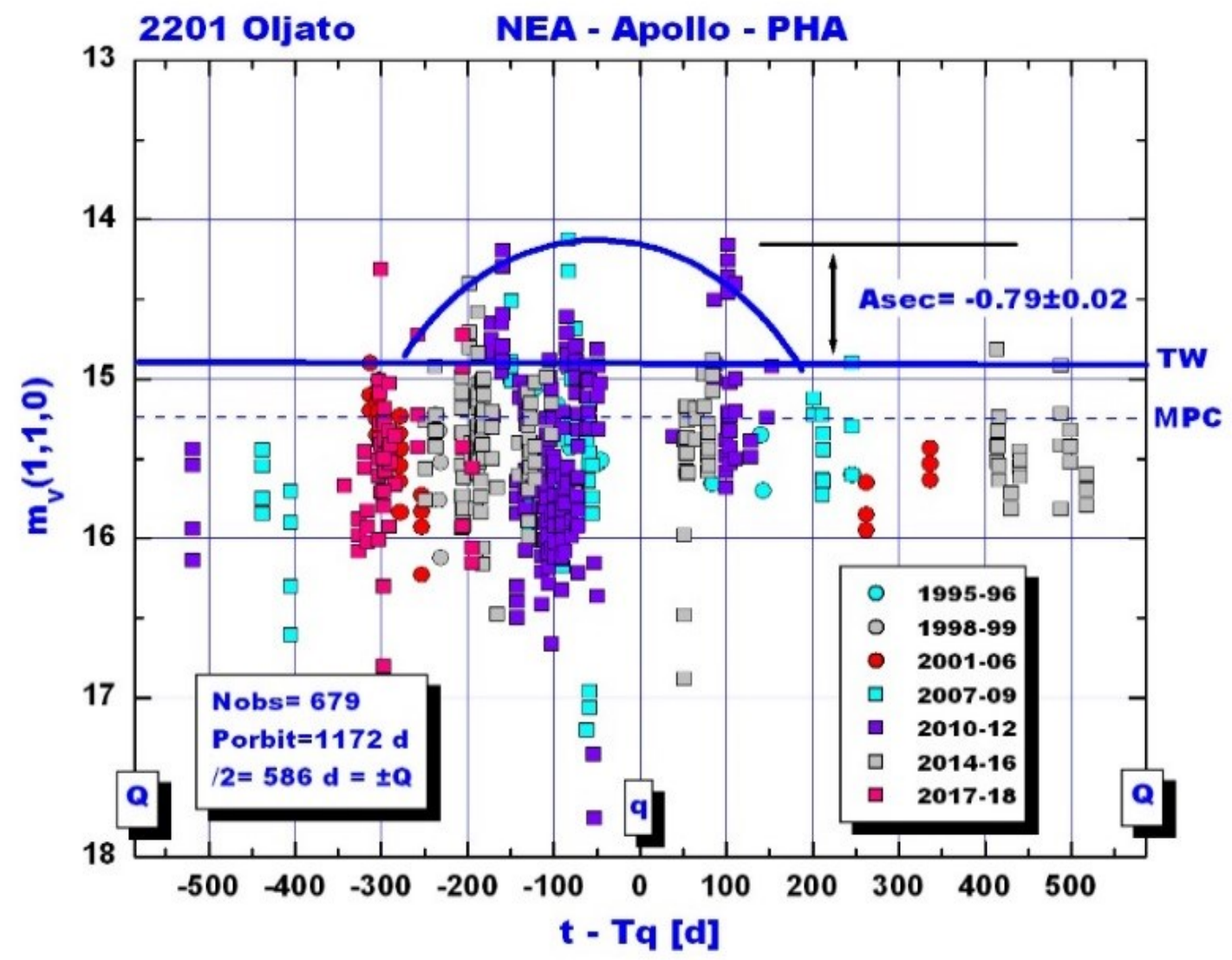

**Figure 6.** SLC of Oljato showing low level cometary activity in 4 different apparitions and strongly suggesting sublimating activity. This plot and Figure 5 show a key element of our reduction procedure. Both plots, phase and SLC, have to agree on the value of the absolute magnitude. This is the most secure way to derive an absolute magnitude because it uses information gathered from two independent phase spaces and the plot covers from –Q to +Q, giving the whole picture of the orbit (see also Appendix 1).

## 4. Components of the complex

As discussed in the introduction, the TC was formed due to the fragmentation of a common ancestor (a huge comet of about 100 km in diameter). This disintegration generated both large and small objects. Among the former we can mention some traditional comets (such as 2P), several relatively large asteroids (such as Oljato), and many small asteroids (such as that of Tunguska and most likely also that of Chelyabinsk). In this regard, it should be remembered that some of these objects could not directly come from the fragmentation of the original parent body but from the further fragmentation of a secondary body originating from a previous breakdown of the great ancestor. In these case groups of objects with orbital parameters very similar to each other are often obtained, also as regards the longitude of the perihelion which is the parameter that shows the greatest variation among the objects of the complex (see Table 5).

Small objects (meteoroids, with size less than 1 m) are generally made up of particles released by comets that are large enough not to be affected by non-gravitational radiative effects (mainly radiation pressure and Poynting-Robertson effect) that are able to remove the smallest particles away from the complex. These large particles tend to arrange themselves approximately along the orbit of parent body, forming the so-called dust trails. When the Earth intersects these trails, the latter give rise to meteor showers.

In this section we will examine more in detail some of these components of the complex, starting with the two important groups Hephaistos and 169/P.

### 4.1 The Hephaistos's Group

Table 6 lists all the members of this group. The only important difference between (2212) Hephaistos and 2P is in the value of $\varpi$. Since Hephaistos has a very low D parameter and is also an active asteroid (Figure A2-3), it clearly belongs to the TC. All the members of this sub-group have been identified as active asteroids, except for (382395) that does not have enough observations (N) to derive a light curve.

Galibina and Kastel (1982) did a numerical integration of the orbits of Hephaistos and 2P, finding that a separation probably occurred in the remote past.

**Table 6**. Orbital elements of the Hephaistos's Group. The symbols are the same as in Table1. Notice that all the members of the group, except for (382395), are active asteroids.

| Object | Nobs | Act. Stat. | A [au] | e | i [°] | ϖ [°] | q [au] | Asec [mag] | $\Delta t_A$ [d] | $m_v(1,1,0)$ | Hv |
|---|---|---|---|---|---|---|---|---|---|---|---|
| (2212) Hephaistos | 2505 | + | 2.16 | 0.84 | 11.6 | 237 | 0.35 | 0.60 | 300 | 12.7 | 13.3 |
| (4486) Mithra | 763 | + | 2.20 | 0.66 | 3.0 | 251 | 0.74 | 0.90 | 450 | 15.2 | 15.5 |
| (30825) | 1217 | + | 2.44 | 0.68 | 8.7 | 243 | 0.78 | 0.45 | 100 | 14.9 | 14.7 |
| (85182) | 351 | + | 2.22 | 0.78 | 3.1 | 223 | 0.50 | 0.55 | 220 | 16.9 | 17.1 |
| (106538) | 358 | + | 2.42 | 0.76 | 10.3 | 205 | 0.57 | 0.70 | 290 | 15.9 | 16.2 |
| (192642) | 560 | + | 2.64 | 0.77 | 6.8 | 250 | 0.61 | 0.60 | 50 | 15.5 | 16.3 |
| (382395) | 85 | N | 2.11 | 0.76 | 11.6 | 243 | 0.50 | ---- | ---- | ---- | 16.2 |
| (2001) QE34 | 300 | + | 2.16 | 0.73 | 5.6 | 233 | 0.57 | 0.45 | 40 | 18.3 | 19.0 |
| D/1766 G1 (Helfenzrieder)[a] | ---- | + | 2.66 | 0.85 | 7.9 | 255 | 0.41 | ---- | ---- | ---- | ---- |
| C/1833 S1[b] | ---- | + | ---- | 1.0 [b] | 7.3 | 226 | 0.46 | ---- | ---- | ---- | ---- |

*Notes*

(a) D/1766 G1 (Helfenzrieder) disintegrated and it is no longer visible. Its orbital parameters are consistent with TC membership.

(b) C/1833 S1 is listed in the MPC database as parabolic with e = 1.0, however its small inclination, i = 7.3°, suggests this is a short period comet that belongs to the TC, not an Oort Cloud object (see text, Section 1).

## 4.2 The 169P/NEAT Group

A new hitherto unknown member is comet 169P/NEAT which has its own group (see Table 7) that fulfill the modified D criterion, even if their perihelion longitude $\varpi = 34°$, is far away from the mean value $< \varpi > = 161°$. We include these objects as members of the TC for the reason discussed in Sect. 2 in the particular case of Adonis, and also because three of them show cometary activity. 169P/NEAT is a *bona fide* comet of large amplitude of the SLC ($A_{sec} = -7.2 \pm 0.2$) equal to the activity of 2P ($A_{sec} = -7.2 \pm 0.2$) (Figure 8). The dispersion of the group in the ϖ parameter is very small (28° to 35°) suggesting that the group was originated by the disruption, occurred not long time ago, of a fragment detached from the TC ancestor in a more remote past (see discussion in Sect. 4).

**Table 7.** The 169P/NEAT Group. The symbols are the same as in Table1.

| Object | Nobs | Act. Stat. | a [au] | e | i [°] | ϖ [°] | q [au] | Asec [mag] | $\Delta t_A$ [d] | $m_v(1,1,0)$ | Hv |
|---|---|---|---|---|---|---|---|---|---|---|---|
| 169P/NEAT | 1250 | + | 2.60 | 0.77 | 11.0 | 34 | 0.60 | -7.2 | 600 | 15.9 | ---- |
| (2101) Adonis | 131 | + | 1.87 | 0.76 | 1.3 | 34 | 0.44 | -2.90 | 290 | 19.0 | 18.8 |
| P/2003 T12 | 287 | + | 2.59 | 0.77 | 11.0 | 35 | 0.60 | -0.75 | 230 | 18.25 | ---- |
| (1995) CS | 14 | N | 1.94 | 0.77 | 2.6 | 28 | 0.44 | ---- | ---- | ---- | ---- |
| (139359) | 300 | + | 2.63 | 0.87 | 5.9 | 26 | 0.34 | -0.35 | 140 | 15.75 | 16.4 |

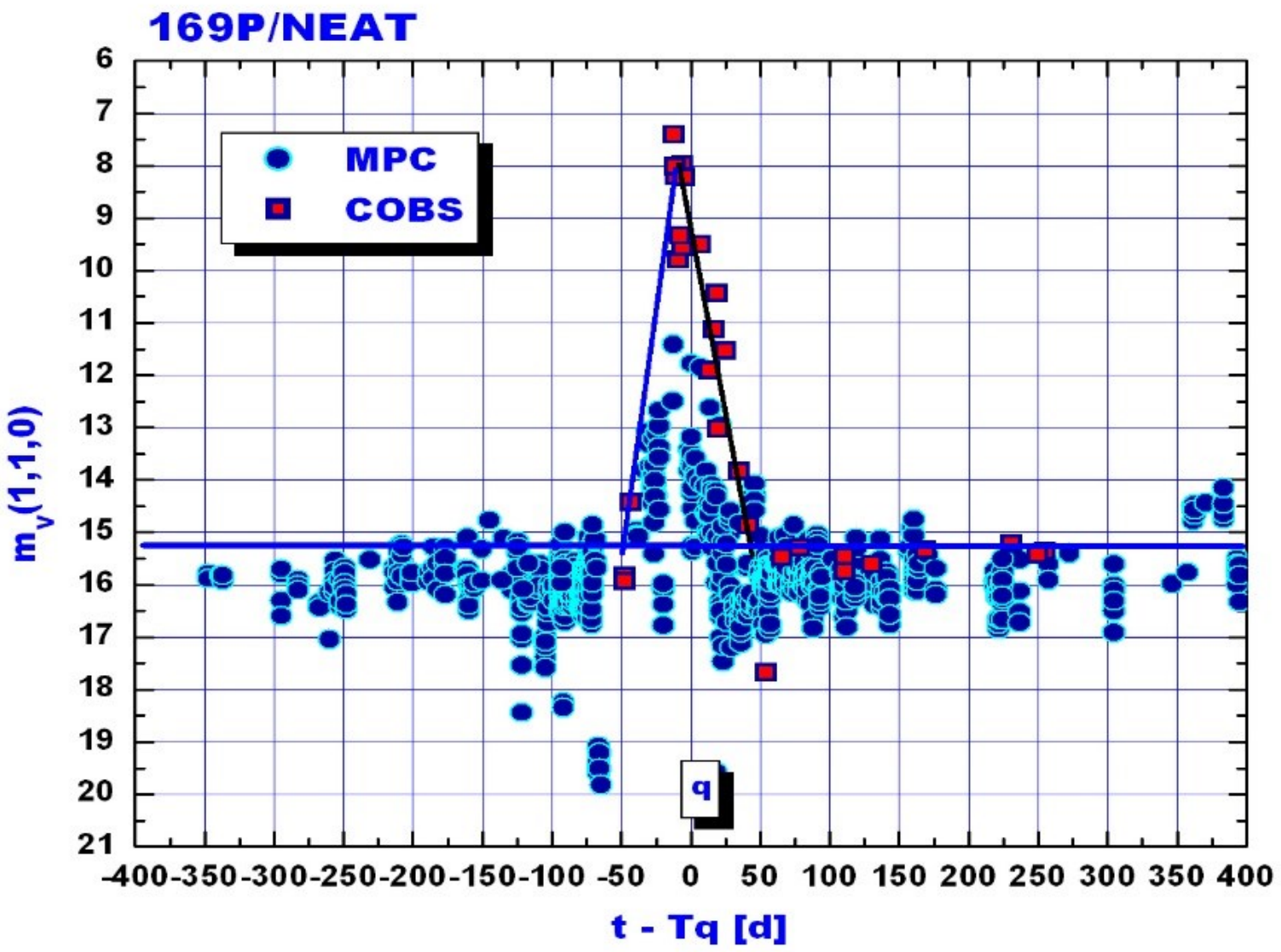

**Figure 7**. SLC of 169P/NEAT, a member of the TC and leader of a group of five active asteroids (Table 7). The SLC is notable for its intensity ($A_{sec} = -7.2 \pm 0.2$ mag), and for its short duration (98 days). The comet rivals in intensity to 2P/Encke ($A_{sec} = -7.2 \pm 0.2$ mag). The COBS (2020) web site gives visual magnitudes (red squares) that agree very well with the MPC data, with no correction at all.

## 4.3 The Tunguska and Chelyabinsk cosmic bodies

On June 30, 1908, a small asteroid entered the Earth's atmosphere exploding 8 km above the ground in an uninhabited area of Russia crossed by the Tunguska river. Fortunately, the explosion did not produce human casualties but caused the felling of millions of trees, devastating an area more than 2000 square kilometers (Gasperini, Bonatti and Longo, 2008).

Unfortunately, the orbital parameters of the so-called Tunguska Cosmic Body (TCB) are not known, but the approximate position of the radiant and the time (day and hour) of the arrival of the object immediately made scholars think that the exploded body belonged to the Taurids and that in particular it came from the comet Encke (Kresak, 1978). On the other hand, subsequent studies have shown that the position of the radiant and the arrival velocity are incompatible with an orbit of a traditional (short or long period) comet, rather indicating orbital characteristics of a Near-Earth Asteroid (NEA) (Andreev, 1990; Sekanina, 1998; Farinella et al., 2001). Although Encke atypical comet has an orbit that closely resembles that of a NEA, a direct origin from that comet was discarded by Sekanina (1998) on the basis of several arguments, partly contested by Asher and Steel (1998).

An important step towards understanding the nature of the TCB has been made thanks to the numerical simulations of Jopek et al. (2008). These authors have chosen a sample of 3311 particles (TCB clones) characterized by radiant coordinates (azimuth, A and height, h) and an arrival velocity (V) consistent with the observations; then, they have extrapolated backwards in time the trajectories of the clones searching for the one with the highest orbital similarity (lowest D) and the minimum orbit intersection distance (MOID) with respect to an object of a selected group of 1340 asteroids and 35 comets (2P included). Jopek et al. (2008) found that the object that best fits the previous requirements is the TCB clone No 2207 (A = 97°, h = 26°, V = 26 km/s), that in the year 932 BC had, with respect to the active asteroid (106538) 2000 WK63 (see Table 2), the minimum value of the D parameter (0.0237) and a MOID equal to 0.01 ua. The latter is low enough to easily become zero, when the non gravitational forces acting on the clones, neglected in the simulations, are taken into account. It is therefore probable that the Tunguska object detached more than 1000 years ago from the asteroid (106538).

The fact that, according to the results of Jopek and colleagues, the TCB appears to come from an asteroid belonging to the TC strongly suggests that the meteoroid is itself a member of that complex. To verify this, we evaluated the orbital parameters of the TCB clone No. 2207, starting from its values of A, h and V (reported above) and using Fig. 2 of Andreev (1990). We thus obtained a = 2.50 ua, e = 0.80 and i = 4.0° and consequently a value D = 0.13; on the other hand, also the longitude of the perihelion, equal to 179°, is not too far from the reference value ($\varpi_R$ = 140° - see above) and all this confirms the belonging to the TC. This suggest that the TCB, although it does not come directly from the Encke comet, should still be associated with this comet which is the progenitor of the TC. In other words, some time ago (of the order of tens of thousands of years) the TCB, asteroid (106538) and comet 2P, and more in general all the members of the TC, were most likely part of a single large original body.

Another hazardous event for our planet occurred on February 15$^{th}$, 2013 when a small asteroid, arriving on the Earth with a low angle of atmospheric entry, exploded at a height of about 30 km over the Russian town of Chelyabinsk. The shock wave due to the explosion broke the glasses of the windows of many houses and about 1,500 people were seriously injured. Furthermore, the shock wave structurally damaged 7200 buildings in six cities across the region.

In Table 8, we report the various sets of the definitive orbital parameters of the Chelyabinsk cosmic body (CCB) taken from the literature. As one can see, the average values of D and ϖ are consistent with the belonging of the small asteroid to the TC.

The different arrival date on Earth with respect the Tunguska object could be simply due to the different values of the longitude of perihelion for the two bodies, in turn due to the their different age (time elapsed from the fragmentations that originated the bodies). The different values of ϖ for the two objects, in fact, imply different orientation of their orbital major semi-axes with respect to the terrestrial one and by consequence different intersection points of these objects with Earth orbit.

With the inclusion of the Tunguska and very probably also of the Chelyabinsk object into the TC, the hazard from this complex has been raised to a new level of concern.

**Table 8**. Orbital parameters of the CCB: for this object D = 0.29 ± 0.06.

| a [au] | e | i [°] | ϖ [°] | D | Authors |
|---|---|---|---|---|---|
| 1,76 | 0,58 | 4,9 | 74,7 | 0,27 | (1) |
| 1,72 | 0,57 | 5,0 | 74,2 | 0,28 | (2) |
| 1,47 | 0,52 | 4,6 | 63,1 | 0,37 | (3) |
| 1,62 | 0,53 | 4,0 | 76,2 | 0,33 | (4) |
| 1,88 | 0,61 | 5,9 | 75,3 | 0,23 | (5) |
| 1,69 | 0,56 | 4,9 | 72,7 | 0,29 | Average |

*Notes*
(1): Popova et al. (2013); (2): Borovicka et al. (2013). (3): Proud (2013); 4): de la Fuente Marcos & de la Fuente Marcos (2014); (5): Emel'yanenko et al. (2014).

## 4.4. Analysis of the IAU's Meteor Database

As mentioned in the Sect. 4, in addition to relatively large bodies (such as traditional comets and both large and small asteroids), much smaller objects also belong to the TC, generally produced by sublimation of cometary ices and by collisions between larger bodies. When they enter the terrestrial atmosphere, they often give rise to meteor showers.

The IAU maintains a database of meteor showers and their radiants, and the data is of interest to define the extent of these component of TC consisting of such very small objects (Jopek and Kanuchova, 2017). Meteor showers have always been associated with comets. In reality, in some cases a shower has been associated with an object of high albedo and therefore with a presumed predominantly rocky surface composition (Dumitru et al., 2017). However, as will be discussed in detail in Sect. 7 in the case of the asteroid (2201) Oljato, a high albedo does not necessarily rule out an inherently cometary origin of the body. So the correspondence between showers and comets continues to hold true.

Figure 8 shows that 30 showers belong to the TC, being mostly contained inside the Taurus Constellation, with two branches parallel to the Ecliptic, the North and the South branch. The Daytime beta-Taurids peak on June 24th while the North and South Taurids peak on October 28th and are dynamically related to a broader complex of 30 showers occurring from June 7th to December 2nd.

The IAU database also gives some parent objects for the showers: 2P (active object, cited 10 times), 2004 TG10 (active, 7 times), 2002 UK11 (without photometric data, 2 times), 2003 WP21 (no data, 1 time), (380455) 2003 UL3 (active, 1 time), 2012 UR158 (no data, 1 time). All have been included in Table 1 except for 2002 UK11 that does not satisfy the modified D criterion ($D = 0.36$).

Due to the above discussed strong correlation between showers and comets (active, dormant or extinct), the presence of many showers belonging to the TC indicates that this complex is not only abundantly populated by comets (67%) but also by a lot of their smaller debris.

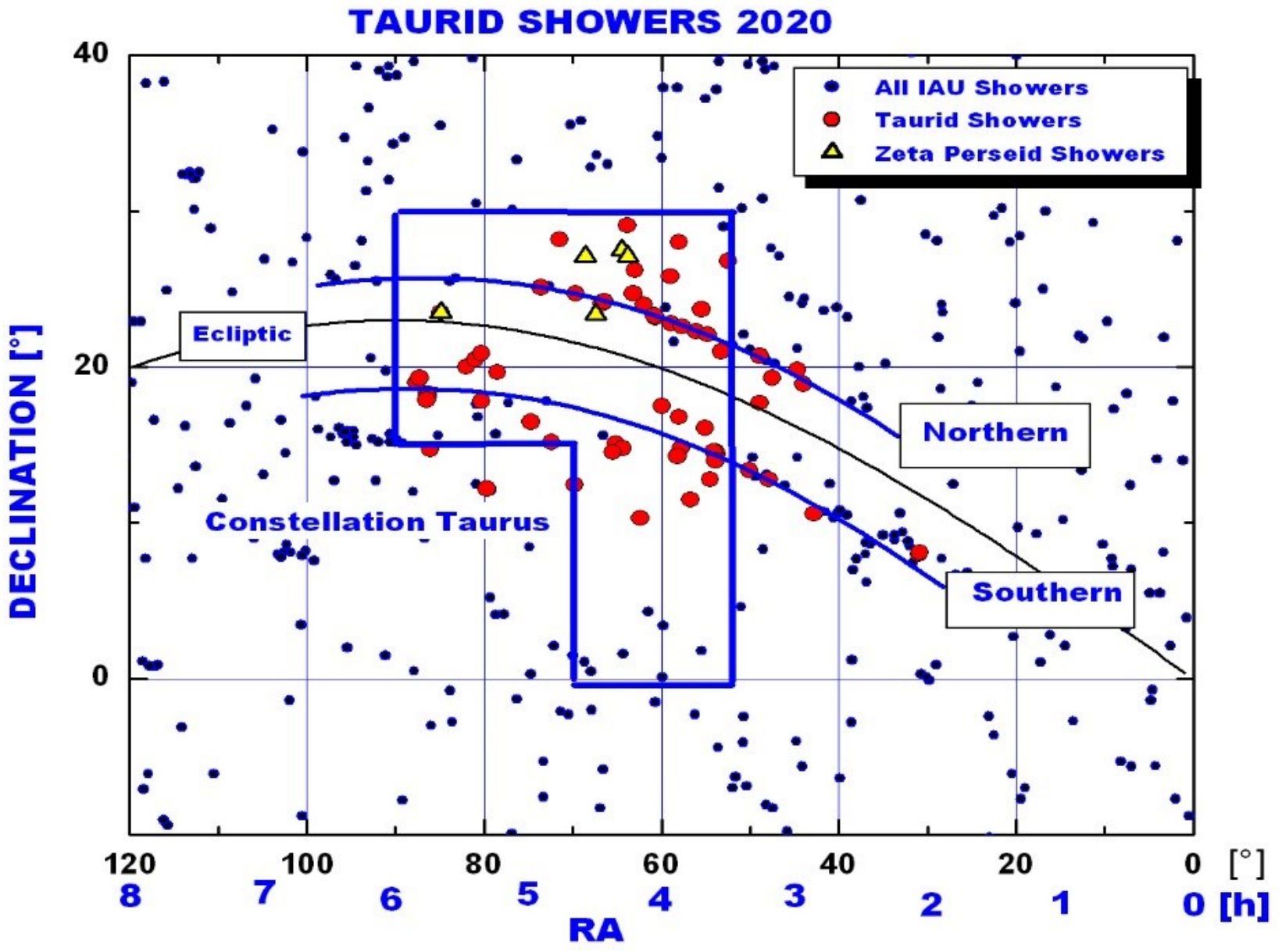

**Figure 8.** Distribution of the Taurid's showers radiants (red circles) in the sky. They lie on two branches, the North and the South symmetrically distributed with respect to the Ecliptic. The diagram shows that the Zeta Perseids shower (yellow triangles) belongs to the TC.

## 5. How odd is 2P/Encke?

2P is, among all the objects classified as comets, one of the oddest according to the following facts.

(1) The perihelion distance of 2P is q = 0.34 au, while its aphelion distance is 4.09 au, thus the comet is completely detached from Jupiter.

(2) It is, among all the bodies with typical NEA orbits, the one exhibiting the most robust cometary activity of amplitude $A_{sec}(q) = -7.2 \pm 0.2$ magnitudes, indistinguishable from a *bona fide* comet. Thus the question is, how did it get to the region where we see it today? Or alternatively, has it always had this kind of orbit?

(3) The SLC of 2P (Figure 9) suggests the presence of two active regions on its surface: Source 1 at the North pole that points to the Sun at perihelion and shows an amplitude Asec(q) = −7.2 ± 0.2 magnitudes; and Source 2 at the South pole that points to the Sun at aphelion with Asec(Q) = −3.0 ± 0.2 magnitudes (Sekanina, 1979; Ferrín, 2008). Source 1 does not see the Sun near aphelion and Source 2 does not see the Sun near perihelion. This situation is very odd and we do not know of any other comet with this particular configuration. The object is elongated with the poles lying almost on the ecliptic plane and with one pole pointing to the Sun at perihelion.

(4) The phase coefficient of 2P is $\beta = 0.066 \pm 0.003$ (Ferrín, 2008) very different from the mean value of the TC, $\beta = 0.037 \pm 0.002$ (see Figure 10). The surface structure of an active *bona fide* comet might be very different from that of an evolved active asteroid. This is the best evidence that the TC harbors objects of different composition and spectral classes (see also Sect. 7). Matlovic et al. (2017) analyzed 33 Taurid meteor spectra finding that the spectral and physical characteristics point towards cometary origin with highly heterogeneous content, confirming our result.

(5) Levison et al. (2006) published a work on the [quoted] *"unusual"* orbit of 2P. They tried to explain the origin of the comet from beyond the orbit of Jupiter by integrating a large number of test particles. They found that it takes roughly 200 times longer to evolve onto an orbit like the present one than the typical cometary physical lifetime. Actually, according to previous dynamical calculation carried out by Pittich et al. (2004), a much smaller timescale is obtained, if non-gravitational forces are included in the calculation. However, Levinson et al. (2006) (who did not include non-gravitational forces in their calculation), conclude that the non-gravitational forces used by Pittich et al. (2004) were unrealistically large and held constant. So the problem of the time required for 2P to evolve onto the present orbit, remains. To solve this problem Levinson et al. (2006) propose that: (a) 2P became dormant soon after it was kicked inwardly by Jupiter; (b) it spent a significant amount of time inactive while rattling around the inner Solar System; or (c) it only became active again as the $\nu 6$ secular resonance drove down its perihelion distance.

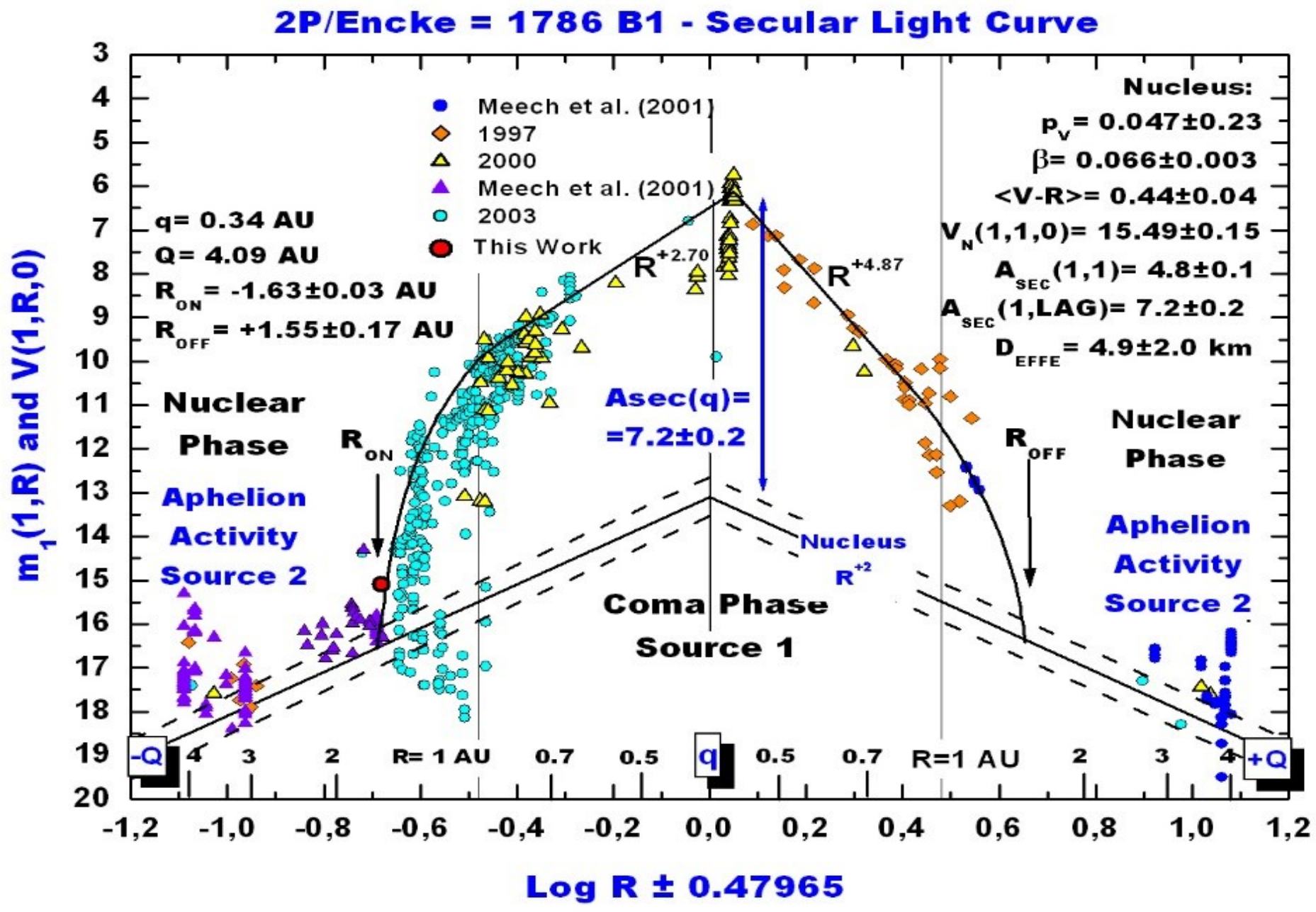

**Figure 9**. SLC of comet 2P, the parent of the TC (Ferrín, 2008). This comet exhibits two sources of activity: Source 1, centered at perihelion, and Source 2, centered at aphelion (data points for R < −3.0 au and R > +3.0 au). Source 1 turns on at $R_{1on}$ = −1.63 au from the Sun (pre-perihelion) and turns off at $R_{1off}$ = +1.55 au (post-perihelion). So, if the 2P orbit were circularized with a radius equal to aphelion distance (4.09 au), then Source 1 would be inactive ($A_{sec1}$ = 0.0 mag), while Source 2 would still be active ($A_{sec2}$ = −3.0 mag). Other information on 2P relevant to this investigation is found in Sekanina (1979), and Ferrín (2008).

(6) The ν6 secular resonance is located at 2.05 au from the Sun (Scholl and Froeschle, 1991) and thus it is well inside the region of activity of Source 2. Our SLC constrains the region of no activity to locations beyond 5 au where it could reside without aging.

(7) By comparing 2P with other comets presented in the *Atlas of Secular Light Curves of Comets* (Ferrín, 2010), it can be concluded that this object exhibits a quite regular activity showing no characteristics that could suggest a period of inactivity. Thus we do not find photometric evidence to support conclusions (a) to (c) above. In fact, there is no physical way in which a comet can rattle around the inner Solar System without exhibiting cometary activity, unless if it were very old and covered with a thick layer of dust. However this is not the case based on its robust activity.

In conclusion, any satisfactory explanation of a non catastrophic origin of 2P lacks in the current literature. The hypothesis of the fragmentation of a giant comet originally proposed by Napier and Clube (1984) solves this conundrum in an elegant way. According with this hypothesis, the progenitor of 2P was a large object placed on a normal short-period orbit coming from the Kuiper Belt. Upon arrival in the inner Solar System, a series of fragmentations, starting from 20 thousand years ago, produced the present object 2P (together with a large number of other bodies), placing it on an orbit drastically different from that of the progenitor body and similar to those of the other NEAs generated by the series of catastrophic event.

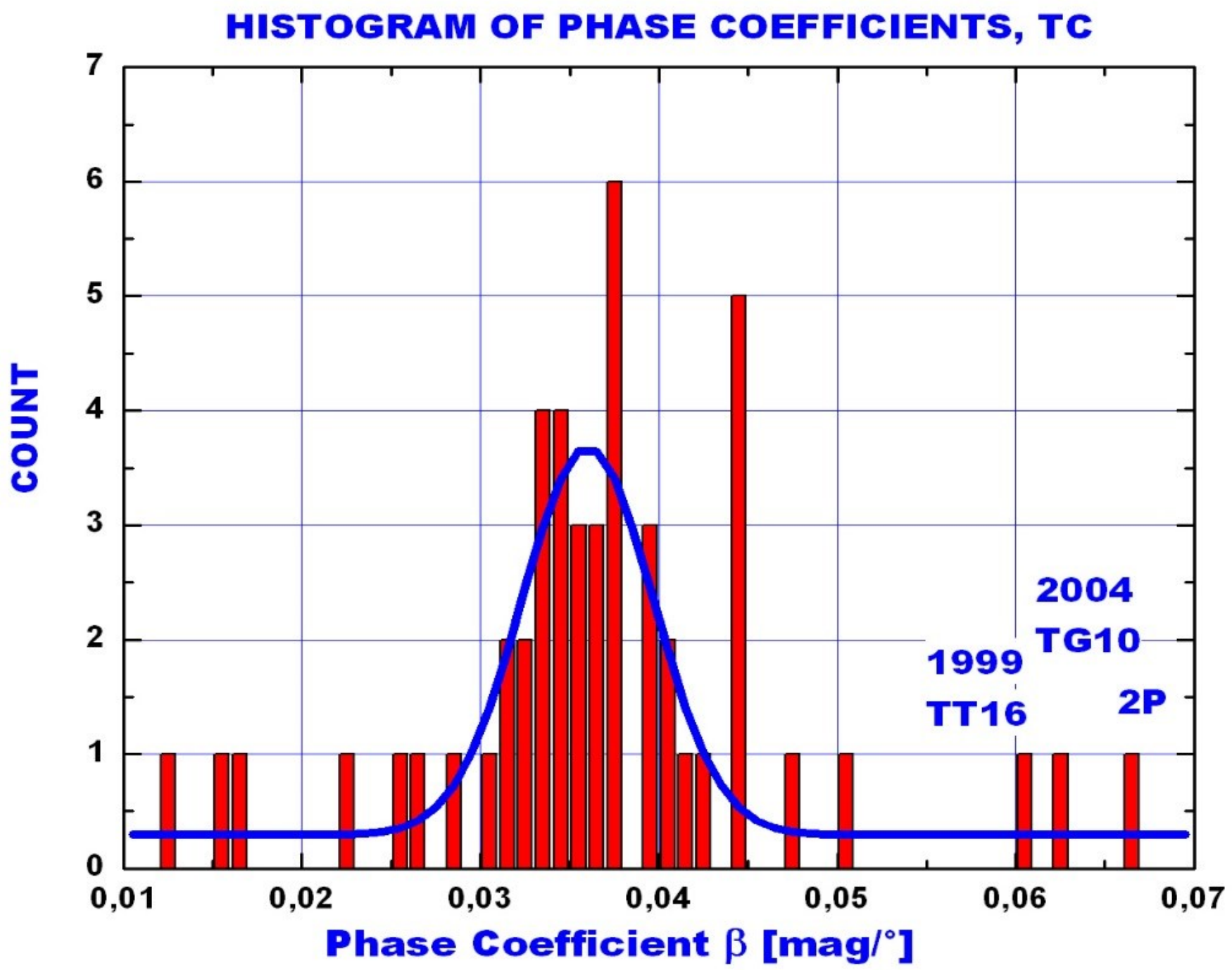

**Figure 10.** Histogram of phase coefficients of the TC objects. It shows that distribution is not homogeneous, with outliers beyond the gaussian fit. For example 2P, 2004 TG10 and 1999 TT16 have values $\beta \sim 0.06$. Another group has values around 0.015, while for the rest of the TC $\beta = 0.035 \pm 0.006$ (Table 1). 2004 TG10 has been proposed as a precursor of 2P (Jenniskens and Jenniskens, 2006) and, after 2P, it is the most cited parent in the IAU Meteor database.

## 6. Backward integration of the diameter

According to Napier and Clube (1984) the spontaneous fragmentation of a large progenitor originated the whole TC. Actually, the fragmentation of a cometary nucleus is a frequent and well documented event (we remind the very recent example of the comet C/2019 Y4 (ATLAS), broken in more than 12 pieces at the end of March 2020) that can also be recurrent for the same object (Boehnhardt, 2004). Even if the tidal disruption is the best studied mechanism of fragmentation (Sekanina, 1994), other mechanisms like thermal stress, rotational instability, or collisions have been proposed (Sosa and Fernández, 2015). Another possibility is that the original nucleus splits up into two or more pieces due to the jet action of outgassing from sublimating ices. Because such venting is usually not evenly dispersed across the comet, it might cause the breakup.

In the framework of the fragmentation hypothesis, an important question concerns the size of the parent body that Napier and Clube (1984) and Napier (2010) estimate around 100 km (in diameter). Due to the large number of objects of our sample, we can approximately evaluate, with reasonable assumptions on their albedo, the total volume of the present members of the TC; this volume represents a lower limit for the original one of the parent body of the TC because we ignore the mass of debris associated with the complex, as evidenced by the meteor showers.

The absolute magnitude $H$ of an object can be used to evaluate its diameter, and by consequence its volume, if a geometric albedo ($p_V$) is adopted, by means of the equation (Lamy et al., 2004):

$$d = \frac{1336}{\sqrt{p_V}}\, 10^{-\,0.2\,H}$$

where $d$ (in km) is the diameter of the object. In this equation we have used the absolute magnitude obtained by means of the SLC method ($H = m_V(1,1,0)$) in the present or in previous works of one of us (IF); when not available, we have used the $H$ magnitude listed in the MPC database with a corrective term (−0.6), which represents the average difference between our measurements of $H$ and those reported in the MPC catalogue. This difference is due to the fact that the MPC catalogue has a tendency to give fainter absolute magnitudes because they use the mean value of the data, while in the SLC formalism we use the envelope.

By summing the volumes of each member (comets excluded) obtained for a fixed value of albedo, we calculated the two limit values of the total volume of our sample in the two cases: a pure rocky composition (similar to that of a S-type asteroid), dominated by silicate materials, with $p_V = 0.20$ (DeMeo and Carry, 2013); and a pure non-rocky (typically comet-like) composition, dominated by hydrocarbons and ices, with $p_V = 0.04$ (Fernandez et al., 2000; Kokotanekova et al., 2017). Starting from these limit volumes, a more reliable value of the total volume was obtained, as a weighted average of the rocky and non-rocky components of the complex, with the weights given by the percentage by volume of the rocky (22% silicates) and non-rocky (54% hydrocarbons, 4% Fe-sulphides, and 20% ices) components of the nucleus of comet 67/P Churyumov-Gerasimenko (Fulle et al., 2017), assumed as term of reference for our calculations. In this way we have found for the members of the TC that are not catalogued as comets a total volume 8500 km$^3$.

As far as comets are concerned, the diameter of P/2003 T12 has been determined, as before, starting from its $H$ magnitude (determined by us) and assuming $p_V = 0.04$; the diameters of 2P and 169P/NEAT were directly taken from Lamy et al. (2004) and Kasuga, Balam and Wiegert (2010), respectively, while for comet D/1766 G1, since no data are available, the diameter was assumed equal to the average diameter of the Jupiter Family comet nuclei (Fernandez et al., 1999). For these four bodies we have found a total volume of 140 $km^3$, which results in a total volume of our whole sample equal to 8600 $km^3$. We expect that this total volume is close to the true one, since most of the largest TC members are present in our sample and any member left will be faint and thus small. This total volume implies a minimum diameter D′ = 25 km of the parent body. This diameter refers to the time of the event that originated the TC, that we conservatively assume occurred 20,000 years ago (see Section 1).

An improved evaluation of this lower limit for the diameter of ancestor of the TC can be obtained evaluating the mass lost due ice sublimation from the object. In order to integrate backwards 20,000 years the radius of the TC ancestor, as a first step we have neglected the fragmentation of the nucleus, considering the ice sublimation from a single body and using an important property of sublimating comets: the thickness of the layer lost per apparition is approximately constant at every return. This can be inferred from the energy conservation equation. The energy captured from the Sun depends on the cross-section of the nucleus, $\pi r_N^2$ (where $r_N$ is the nuclear radius), on the Bond albedo, $A_B$, and on the solar constant, S. The energy conservation equation can be written as follows:

$$(1 - A_B)\, S\, \pi\, r_N^2 = \varepsilon_{IR}\, \sigma\, T^4 + K_1\, L\, 4\, \pi\, r_N^2\, \Delta r_N + K_2\, \partial T/\partial x\ .$$

Here $\varepsilon_{IR}$ is the emissivity of the nucleus in the infrared, T is the temperature, $K_1$ and $K_2$ are constants, σ is the Stefan–Boltzmann constant, L is the latent heat of sublimation, $\Delta r_N$ is the thickness of the layer removed, and x is the depth below the surface. The term on the left is the energy captured from the Sun. The first term on the right is the radiated energy, the second is the energy lost by sublimation, and the third is the energy conducted into the nucleus. The first and third terms on the right-hand side are small in comparison with the second term because at large distances the temperature and conductivity are very low. The second term dominates near the Sun at perihelion. So, as a first approximation:

$$(1 - A_B)\, S\, \pi\, r_N^2 \sim K_1\, L\, 4\, \pi\, r_N^2\, \Delta r_N$$

$$\Delta r_N \sim (1 - A_B)\, S /(4\, K_1\, L)\ .$$

We find that the thickness of the layer removed by apparition, $\Delta r_N$, should be approximately constant as a function of time.

Ferrín (2014) has determined that 2P has $r_N / \Delta r_N = 8580$ returns left, thus, adopting the radius of 2P ($r_N$= 2400 m), reported by Lamy et al. (2004), one has $\Delta r_N \sim 0.28$ m per apparition. Due to the orbital period of 3.30 yr, in 20,000 yr the comet accomplished 6060 returns, losing an equivalent thickness of 6060 × 0.28 m = 1700 m. Now, we can reasonably assume that the equivalent thickness lost by the parent body was about the same thickness lost by 2P, so that, for the original radius of the parent body, we have r’ = r + Δr = 12500 m +1700 m = 14.200 m, that implies D′ ≥ 28 km. If we assume a density of 533 $kg/m^3$, equal to that of the well-studied comet 67/P Churyumov-Gerasimenko (Patzold et al., 2016), we find an original mass of the parent body $M > 6 \times 10^{15}$ kg.

All this neglects fragmentation. If we consider this process, we have that the previous estimate is very conservative since the fragmentation increases the mass loss by sublimation (due to the increased surface); in addition fragmentation involves a further loss of mass, precisely due to the expulsion of dust subsequently removed from the complex (by radiation pressure and/or Poynting-Robertson effect) or still present in form of unseen particles too small to be detected. It is not easy to evaluate the mass of the dust produced and lost by the fragmentation. Clube and Napier (1984) have estimated it equal to 5 x$10^{17}$ kg, obtaining as a result an original diameter of the TC ancestor of about 100 km. Adopting the same value of the lost mass, but with a more updated density of 533 kg/$m^3$ (see above), we obtain a diameter of 120 km, that confirms the large size of the ancestor.

## 7. Structure of the parent body

An important question to be addressed is the coexistence in the same complex of active and inactive bodies. For example (4197) Morpheus is a typical inactive asteroid, even if its perihelion distance is quite small (q = 0.52 au), while (2201) Oljato (q = 0.62 au) is active. Note that both asteroids present spectral characteristics similar to the S taxonomic complex (JPL Small-Body Database Browser), suggestive of a mostly rocky superficial composition, while the cometary activity of Oljato would indicate an important ice content.

To explain this apparent conundrum one has to think that most probably the parent body of the TC could have had a mixed composition of rocks and volatile substances. It could have been formed beyond the so-called *snow line* (which is now at about 3.5 au from the Sun, but that in the past was closer to our star due to its lower luminosity – Gough, 1981) by heterogeneous coalescence of icy and rocky bodies. The latter were sent in those external regions of the Solar System by the migration of the planet Jupiter (Walsh et al., 2011).

The ancestor of the TC can be imagined as an object with a structure similar to a so called *rubble pile* (Weidenschilling, 1994), that is made by elementary, rocky or carbonaceous blocks, held together inside an icy matrix. Such an object, due its fragmentation, could have originated various objects both relatively large, with the same heterogeneous *rubble pile* structure, and quite small, consisting of the original rocky blocks. Oljato (D ≈ 1.8 km; JPL Small-Body Database Browser) would be an example of *rubble pile* fragment with external blocks mainly of silicate composition and with an icy matrix still sufficient to produce cometary activity. Large inactive objects, such as Morpheus (D ≈ 1.8 km; JPL Small-Body Database Browser), would be similar to Oljato, but their original icy component could be exhaust or sealed in the interior. Smaller inactive objects, such as most probably 2006 SO198 (due to its large magnitude – see Table 1), could be, instead, the original rocky blocks.

In conclusion, as already observed by Napier (2010), the fact that an object shows a spectrum similar to that of main-belt asteroids does not preclude a cometary nature of that object.

It is worthwhile to note that the key mechanism for the formation of such a heterogeneous parent body should consist in the migration of Jupiter described by the *Grand*

*Tack* model (Walsh et al., 2011). This migration, in fact, would have caused the transfer of rocky bodies from their birth places, close to the Sun, to the region beyond the snow line where they could have coalesced with bodies made by volatile materials (carbonaceous/organic and icy).

## 8. Conclusions

The numerical and statistical results of this work are presented in Table 9. Our conclusions can be summarized as follows.

(1) We analyzed more than 140 objects, taken from various sources (see Section 2), and we have verified that 88 bodies satisfy the D criterion. Due to the probabilistic value of this criterion, all the 88 objects identified in the present work must be considered as probable members of the TC, with a degree of reliability of their membership that is the greater the lower their value of D parameter. The most uncertain member of the TC seems to be the Chelyabinks asteroid, whose membership, however, cannot be absolutely ruled out by the analysis of its orbital parameters and their uncertainties. This object deserves great attention since, together with the asteroid of Tunguska, it highlights the danger of catastrophic impacts with the Earth represented by the members of this complex.

(2) Out of the 88 identified objects, 51 have useful light curves. Of these, 17 do not show signs of cometary activity, while 34 (67%) are active. This a quite high percentage when compared with other asteroidal families as for example Themis family where only the 15% of the members show signs of activity (Ferrín et al., 2017).

(3) In addition to the larger objects that are part of our sample of 88 TC members, the complex consists also of a microscopic component made up of small particles that group in streams and which give rise to meteoric showers when they fall to Earth. This component is important because it constitute the part of the entire system that interacts most frequently with our planet. Since meteor showers are in the vast majority of cases associated with comets (active, quiescent or extinct), they demonstrate that the small component of the complex has overall a clear cometary origin, in accordance with what indicated by our analysis of the TC component consisting of larger bodies.

(4) The two previous results gives support to the idea that the TC was originated by the multiple fragmentation of a comet, a celestial catastrophe that took place about ~20,000 years BP (Clube and Napier, 1984).

(5) (2212) Hephaistos and 169P/NEAT are members of the TC and have their own sub-group. Another member of the complex is the object exploded over Tunguska in 1908 and probably also the asteroid fallen on Chelyabinsk in 2013.

(6) The absolute magnitudes found in this work using the SLC formalism are in the mean 0.6 ± 0.7 magnitudes brighter than the MPC absolutes magnitudes. The MPC has a tendency to give fainter absolute magnitudes because they use the mean value of the data, while in the SLC formalism we use the envelope.

(7) We integrated backwards the size of 2P and found the diameter of the parent body at the time of the first fragmentation, 20,000 years ago, $D' \geq 28$ km. When the dust produced by the fragmentations is included, a diameter of about 120 km is obtained.

(8) As clearly indicated by Figure 9, the TC harbors objects of different compositions and spectral classes. This situation can be explained by a rubble-pile texture (Weidenschilling, 1994) of the ancestor of the TC, which, moreover, due to its intrinsic fragility, is a structure more susceptible to episodes of fragmentation than a compact structure. In particular, this composite structure can explain why some TC members, like Oljato, show a cometary activity (and so an important ice content) coupled with a mostly rocky superficial composition, suggested by their spectral type (Popescu et al., 2014). In fact (see also Section 3), the recurrent increase in brightness during the perihelion passages of such rocky objects suggests the presence inside them of ice which, in proximity to the Sun, sublimates and escapes from the body through superficial fractures. This is suggested, at least in the case of Oljato by some hint of gaseous activity (McFadden et al., 1993; A'Hearn et al., 1995) and could be proved by targeted spectral observations near perihelion (aimed i.e. at the detection of absorption features of water ice near 1.5 and 2.0 μm). So far, however, also due to the small size of these objects, such spectra are unfortunately not available.

**Table 9**. Mean parameters concerning the TC deduced in this work

| **Property** | **Nomenclature** | **Value** |
|---|---|---|
| Approved Members of TC (without 2P) | $N_{id}$ | 88 |
| Number of active asteroids (+) | $N_{active}$ | 34 |
| Number of inactive asteroids (-) | $N_{inactiv}$ | 17 |
| Number with Not Enough Data (N) | N | 37 |
| Fraction of Active Asteroids | f(AA) | 67% |
| Mean Duration of Activity | <Duration> | 199±26 d |
| Mean orbital period | $\langle P_{orbit} \rangle$ | 3.20±0.08 y |
| Fraction of active time (Duty Cycle) | DC | 17.0% |
| Mean Ratio <Porbit> / <Duration> | 1/DC | 5.9 |
| Mean amplitude of activity from the SLC | $\langle A_{sec} \rangle$ | -0.63±0.30 mag |
| Mean phase coefficient | $\langle \beta \rangle$ | 0.035±0.006 mag/° |
| Mean Longitude of Perihelion | $\langle \varpi_{TC} \rangle$ | 161° |
| Longitude of perihelion of 2P | $\langle \varpi_{2P} \rangle$ | 161° |
| Orbital period of comet 2P/Encke | $P_{orbit}$ | 3.290 y |
| 7/2 Resonance with Jupiter for comparison | R(7/2) | 3.389 y |
| Equivalent diameter of parent at 20000 BP | D' | ≥28 km |
| Present total volume of the TC members | $V_{TC}$ | 8600 $km^3$ |
| Systematic error of MPC absolute magnitudes | $\langle H_V\text{-}m(1,1,0) \rangle$ | +0.6±0.7 mag |

## ACKNOWLEDGMENTS

The authors acknowledge with thanks the COBS Comet Observations Database, contributed by observers worldwide and used in this research. The same acknowledgement is valid for the MPC observations database, with its hundreds of contributing observers. The authors also thanks Marcella D'Elia for her support in the orbital data recovery from MPC database. Vincenzo Orofino acknowledges the TAsP and Euclid INFN Projects.

## APPENDIX 1 – Data Reduction

In this work we used the astrometric-photometric database of the Minor Planet Center (Holman, 2019). To reduce the data we used the procedure followed in previous papers on the Secular Light Curve (SLC) formalism (Ferrín, 2008; 2014; Ferrín et al., 2017; 2018). In that formalism we adopt the envelope of the dataset as the correct interpretation of the light curve. There are many physical factors that affect comet observations, like twilight, moonlight, haze, cirrus clouds, dirty optics, lack of dark adaptation, excess magnification, and in the case of CCDs, sky background too bright, insufficient time exposure, insufficient CCD aperture, and a too large scale. All these factors decrease the captured photons coming from the object, and the observer makes an error downward, toward fainter magnitudes. There are no corresponding physical effects that could increase the perceived brightness of the object. *Thus the envelope is the correct interpretation of the light curve.* In fact the envelope is flat, while the anti-envelope (the fainter magnitudes of the distribution), is diffuse and irregular.

The envelope represents an ideal observer (vision 20/20), using an ideal telescope and detector, in an ideal atmosphere (pure and transparent).

A plot of reduced magnitude vs time to perihelion shows some vertical dispersion, because the MPC Observations Database is an astrometric database that uses small photometric apertures to extract the flux, producing fainter magnitudes. However *a well-defined envelope* appears in all inactive asteroids. When we take the envelope of the dataset, we also take the envelope of the rotational light curve. The rotational light curve can be modeled as a sine wave. If a sine wave is sampled at equal intervals, the distribution shows maxima at the two extremes (see Figure 2 of Ferrín et al., 2017). The maximum brightness helps to make the envelope sharper.

To make the envelope even sharper, we reduce all the filtered observations to the V-band using the transformation equations of Jordi et al. (2006). In this way we are able to reduce the uncertainty of the envelope to ~±0.2 magnitudes, our detection limit. In Figure A1-1 we show the flatness of the light curve, after the filter correction.

For asteroids with no activity (a bare nucleus), the absolute magnitude is flat and independent of the time or location on the orbit (by definition) (see Figure 3). On the other hand, objects with activity show localized bumps above the envelope (thus negatively enhanced magnitude), many of them preferentially near perihelion (see Figure 4, as well as Figures A2-1 to A2-5 in Appendix 2).

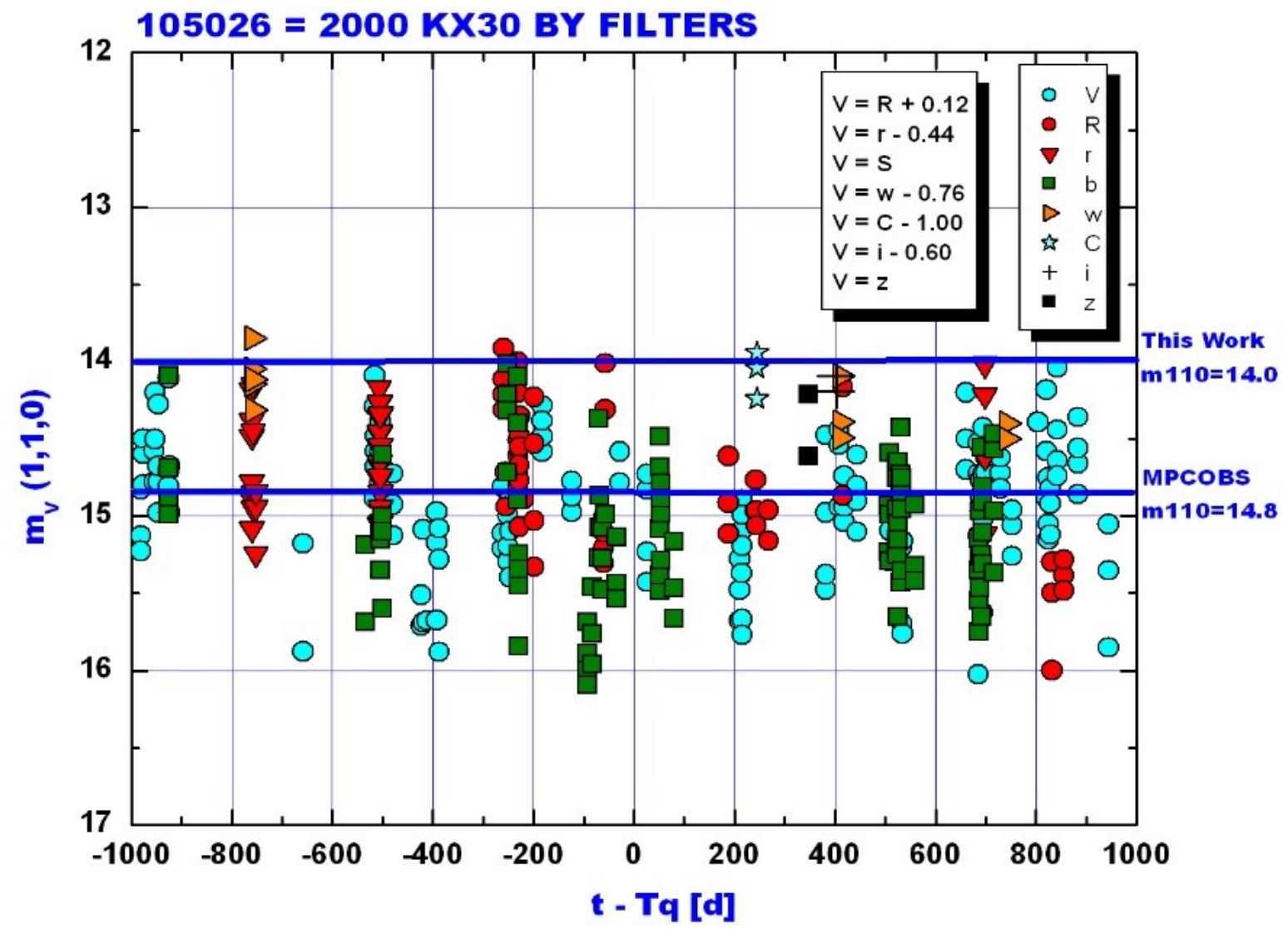

**Figure A1-1.** Effect of reducing the filtered observations of asteroid 2000 KX30 to the V-band. The various filters used in the measurements as well as the transformation equations to the V-band (Jordi et al., 2006) are reported in the boxes near the upper right corner. The scatter of the data points has been reduced to less than ~±0.15 magnitudes. The MPC absolute magnitudes are fainter than observed by 0.8 magnitudes.

The examination of the SLC allows us to declare the activity of an object positive or negative if: (a) the activity appears at the same place of the orbit in different apparitions (see Figure 4); (b) the activity appears in a time span larger than 40 days (our detection limit); and (c) if the (negative) amplitude of the excess brightness is larger than ~0.2 magnitudes (our detection limit after filter correction). Data that shows an excess of magnitude for 1 or 2 days is taken as scatter and not registered as a positive detection.

The quality of our photometric reduction can be assessed for example from Figure 6 with the SLC of asteroid (2201) Oljato. It shows low level cometary activity in 4 apparitions which is the best evidence of sublimating activity, also suggested by two different independent observations performed by McFadden et al. (1993) and by A'Hearn et al. (1995). Also in Figure 4 we present the SLC of 6063 Jason. It shows an enhancement in magnitude near perihelion, at 4 apparitions. Additionally in Figure 7 we show the SLC of comet 169P/Neat, a prominent member of the Taurid Complex. Notice how flat the MPC data are. The absolute magnitude from the MPC data is the same as the absolute magnitude of the COBS data (Comet Observations Database, https://www.cobs.si/). Also, the turn on and turn off days agree very well, -50 days to +48 days with respect perihelion. In Figure 4, the SLC of asteroid 16960 shows low level cometary activity at 5 different apparitions. We conclude that the MPC data base is not perfect, but we have been able to determine the error limits of our detections.

We have also developed a method to determine the absolute magnitude in a more reliable way than other methods. To determine the absolute magnitude we use two plots, the phase plot and the SLC plot. The phase plot shows $m_V(1,1,\alpha)$ the observed magnitude uncorrected for the phase angle, vs phase angle. The SLC plot shows the absolute magnitude $m_V(1,1,0)$ vs time before/after perihelion. The two plots have to agree on $m_V(1,1,0)$. Thus our determination of the absolute magnitude involves two different phase spaces plus the envelope, and thus has to be more reliable than other procedures. For example the absolute magnitudes of the MPC are in the average ~0.6 magnitudes fainter than our results.

# APPENDIX 2 - SLC's of active objects studied in this work

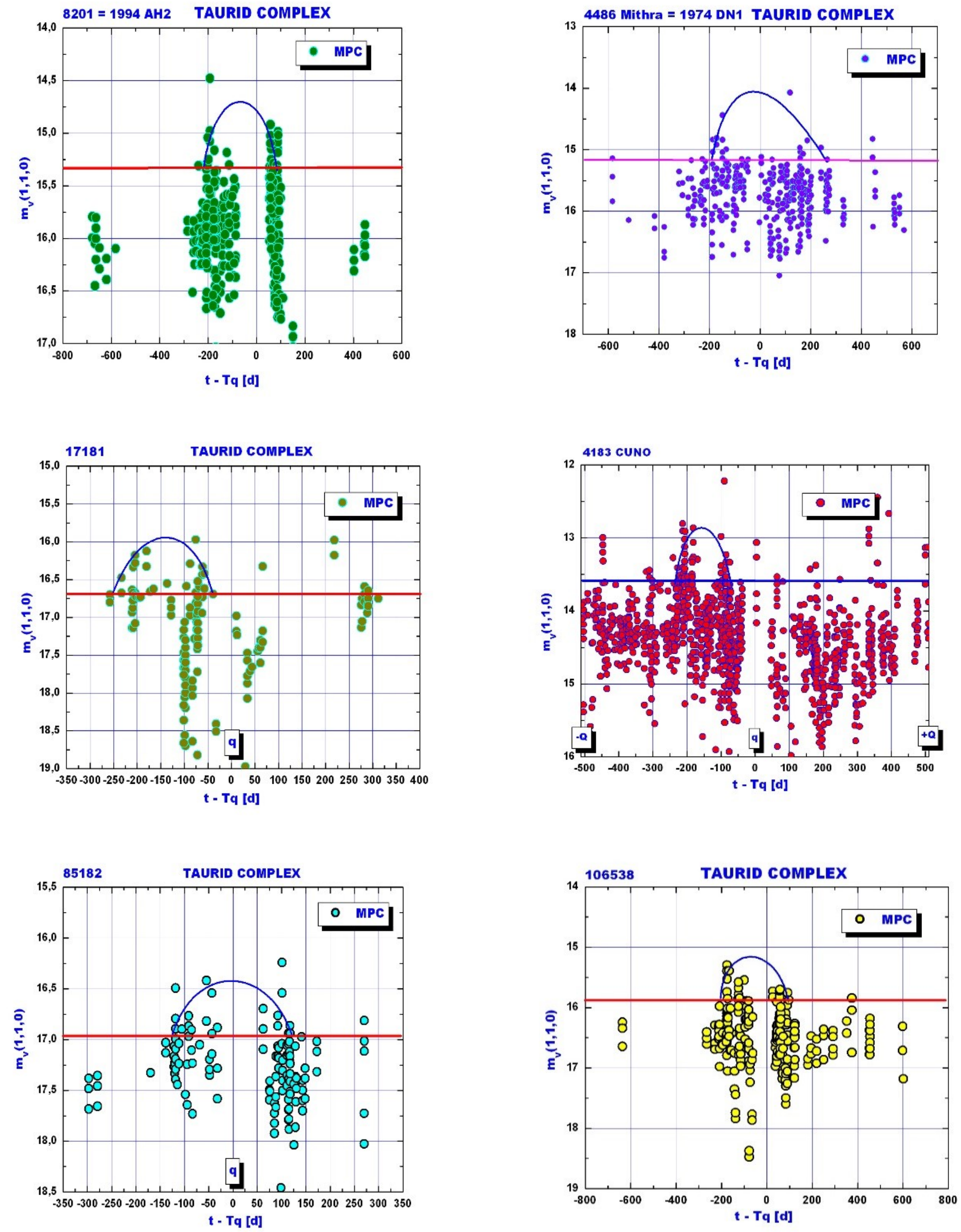

**Figure A2-1**. Additional TC asteroids that exhibit low level cometary activity.

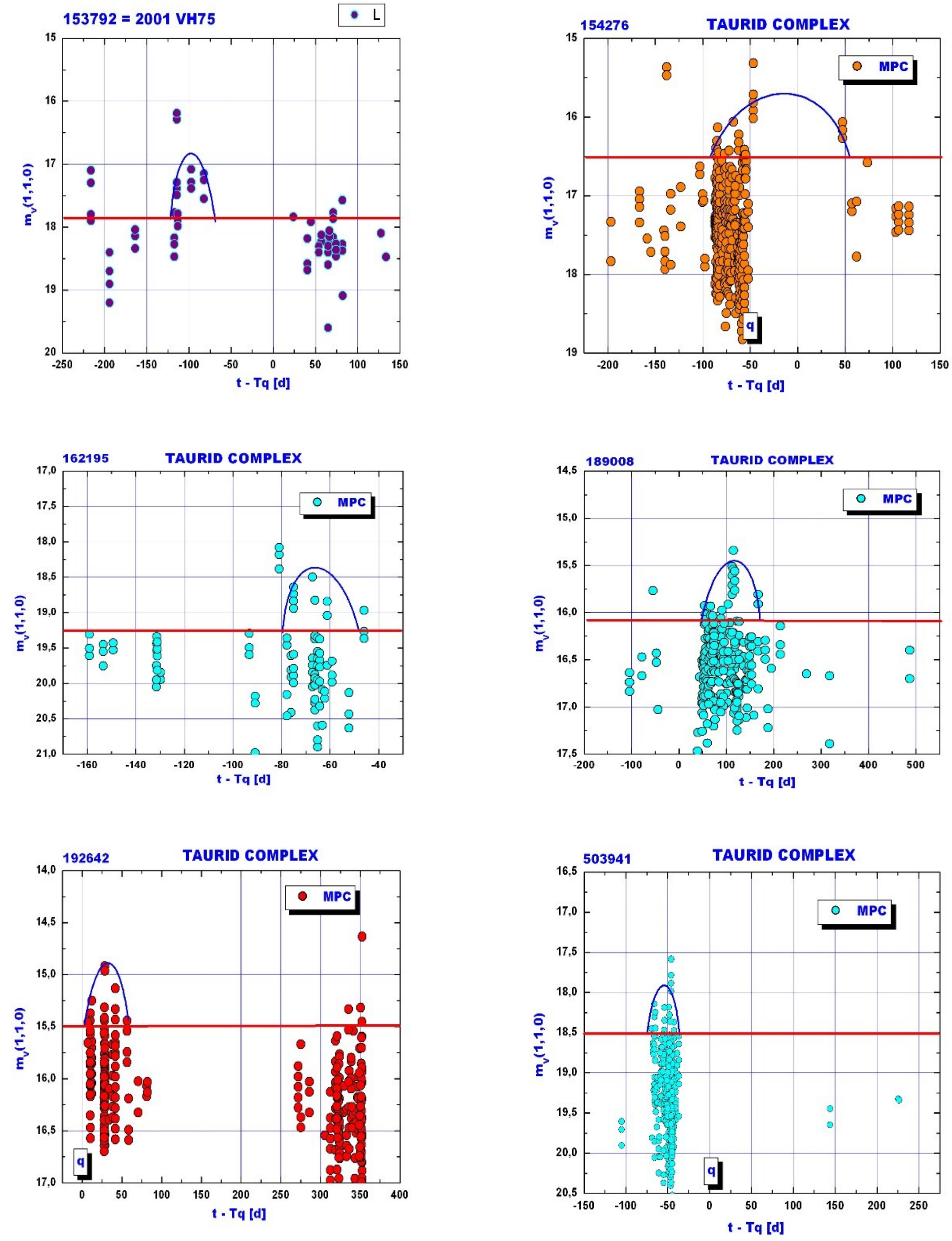

**Figure A2-2**. Additional TC asteroids that exhibit low level cometary activity.

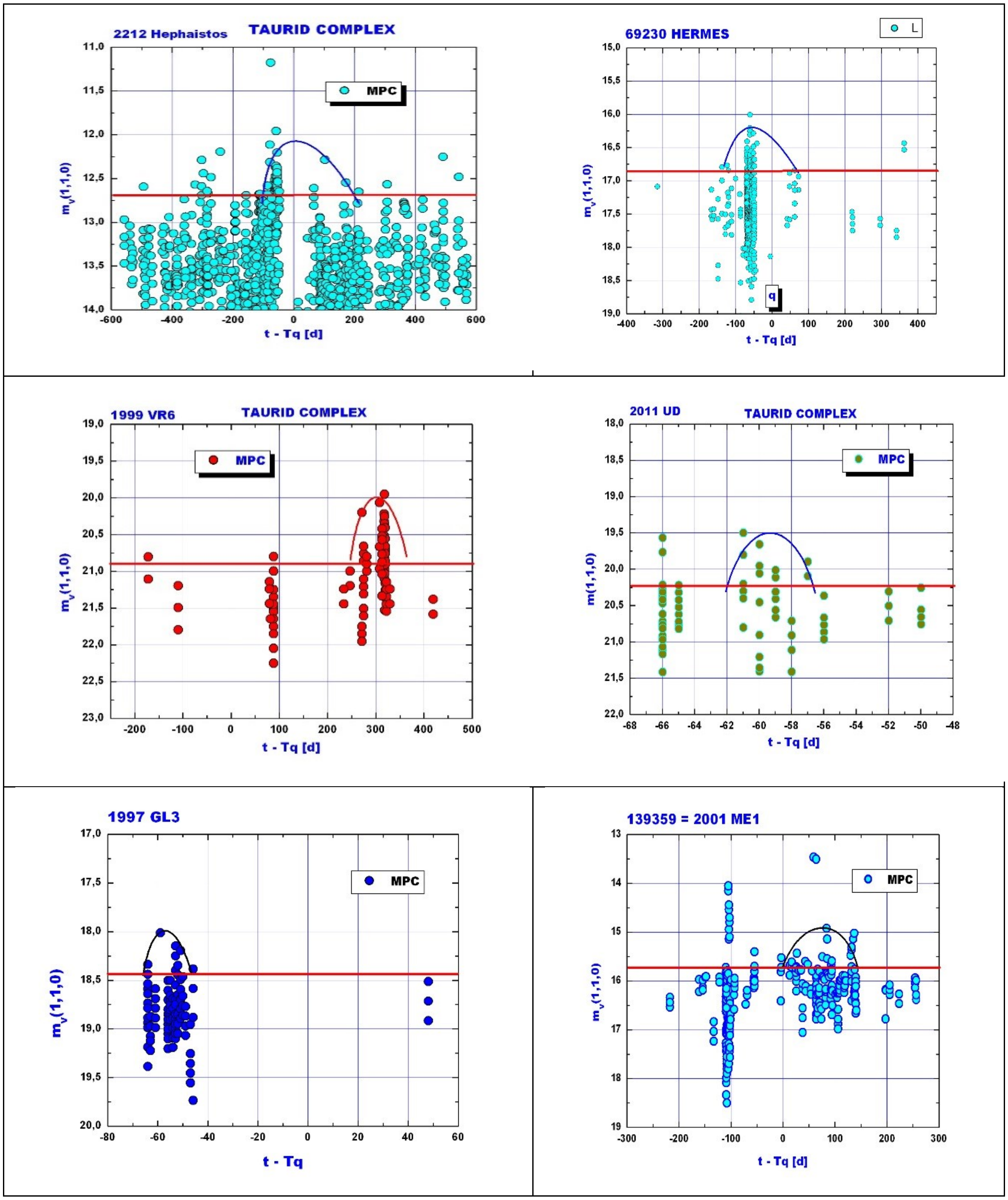

**Figure A2-3.** Additional plots of positive objects.

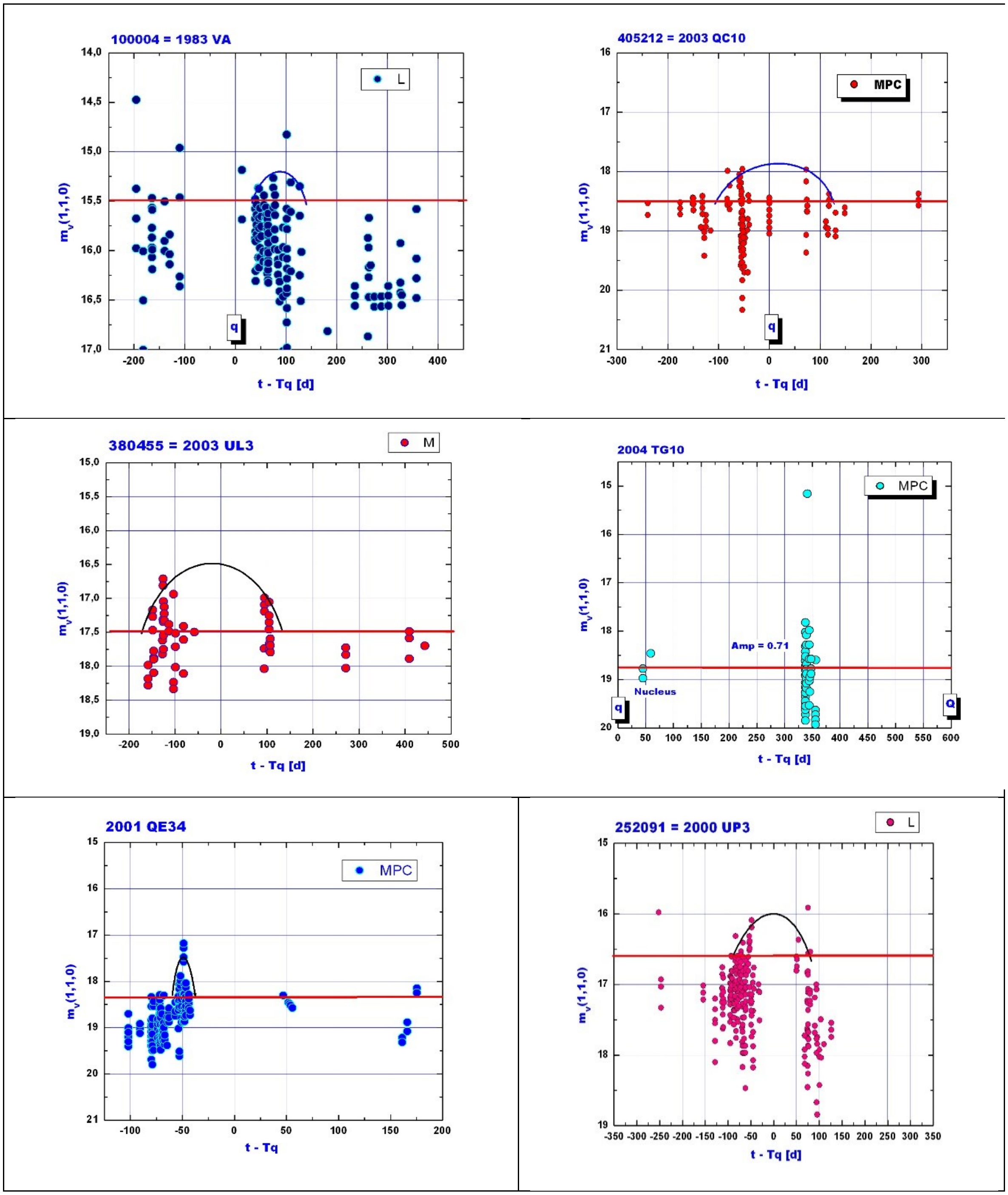

**Figure A2-4**. Additional TC objects positive for cometary activity.

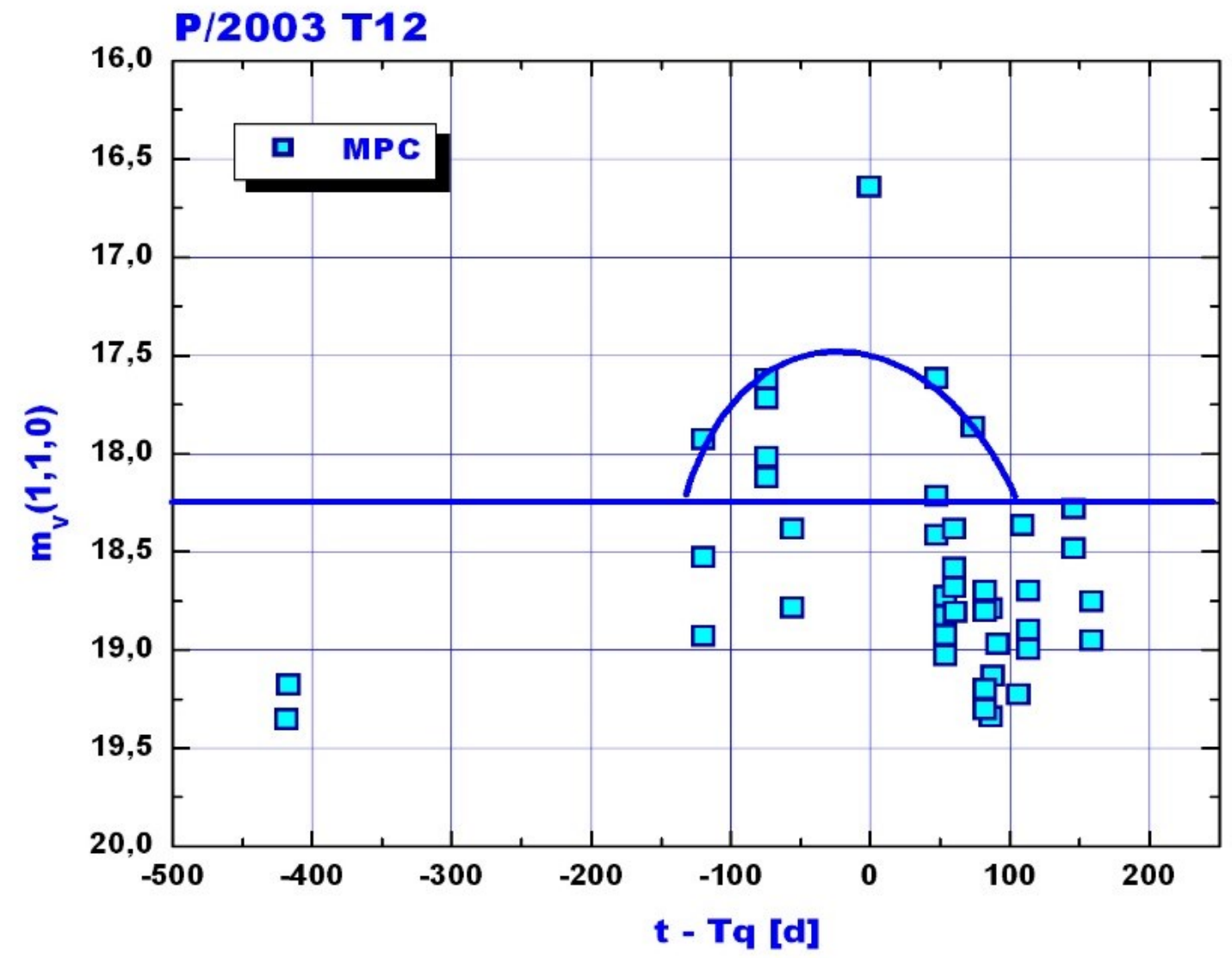

**Figure A2-5.** The SLC of comet P/2003 T12.